# Knowledge-Based Machine Learning Methods for Macromolecular 3D Structure Prediction

By Zhiyong Wang

Submitted to:

Toyota Technological Institute at Chicago

6045 S. Kenwood Ave, Chicago, IL, 60637

For the degree of

Doctor of Philosophy in Computer Science

Thesis Committee:

Jinbo Xu (Thesis Supervisor)

David McAllester

Jie Liang

# KNOWLEDGE-BASED MACHINE LEARNING METHODS FOR MACROMOLECULAR 3D STRUCTURE PREDICTION

by Zhiyong Wang

Submitted to:

Toyota Technological Institute at Chicago

6045 S. Kenwood Ave, Chicago, IL, 60637

For the degree of

Doctor of Philosophy in Computer Science

Thesis Committee:

Jinbo Xu          signature: 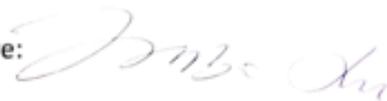   Date: 6/15/2016
(Thesis Supervisor)

David McAllester  signature: 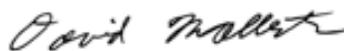   Date: 6/15/2016

Jie Liang         signature: 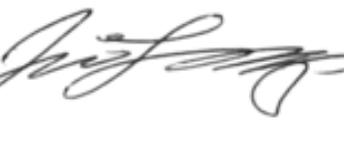   Date: 6/15/2016

# Abstract

Predicting the 3D structure of a macromolecule, such as a protein or an RNA molecule, is ranked top among the most difficult and attractive problems in bioinformatics and computational biology. Its importance comes from the relationship between the 3D structure and the function of a given protein or RNA. 3D structures also help to find the ligands of the protein, which are usually small molecules, a key step in drug design. Unfortunately, there is no shortcut to accurately obtain the 3D structure of a macromolecule. Many physical measurements of macromolecular 3D structures cannot scale up, due to their large labor costs and the requirements for lab conditions.

In recent years, computational methods have made huge progress due to advance in computation speed and machine learning methods. These methods only need the sequence information to predict 3D structures by employing various mathematical models and machine learning methods. The success of computational methods is highly dependent on a large database of the proteins and RNA with known structures.

However, the performance of computational methods are always expected to be improved. There are several reasons for this. First, we are facing, and will continue to face sparseness of data. The number of known 3D structures increased rapidly in the fast few years, but still falls behind the number of sequences. Structure data is much more expensive when compared with sequence data. Secondly, the 3D structure space is too large for our computational capability. The computing speed is not nearly enough to simulate the atom-level fold process when computing the physical energy among all the atoms.

The two obstacles can be removed by knowledge-based methods, which combine knowledge learned from the known structures and biologists' knowledge of the folding process of protein or RNA. In the dissertation, I will present my results in building a knowledge-based method by using machine learning methods to tackle this problem. My methods include the knowledge constraints on intermediate states, which can highly reduce the solution space of a protein or RNA, in turn increasing the efficiency of the structure folding method and improving its accuracy.

# Acknowledgement

There are so many people I am grateful to. I cannot thank too much for those people who gave generous supports to me during the past years.

First, thanks my advisor Professor Jinbo Xu. Without your guidance and great patience, I cannot finish my PhD study. Thanks Professor David McAllester and Professor Jie Liang who advised me on my PhD thesis proposal and defense. Thanks all the professors who taught me during the last 6 years.

Thanks my family who supported me so much.

Thanks my teammates and classmates.

Avleen Bijral

Somaye Hashemifar

Taehwan Kim

Jianzhu Ma

Jian Peng

Xinghua Shi

Siqi Sun

Hao Tang

Qingming Tang

Sheng Wang

Payman Yadollahpour

Feng Zhao

It is my great honor to have the chance to work and study with you together.

# Table of Contents

# Chapter 1. The life of a protein and a RNA

## Genetics Origination

Before we dive into the protein/RNA structure prediction study, we first need to have a map of the positions of protein and RNA in a living organism. Though many biological college text books introduce what is a protein/RNA and how they are produced, employed and decomposed in living cells, they are usually presented with a great detail. The low level detail of biology would bury the high level meanings of their functions. In fact, the high level functions of protein and RNA give a whole picture of the biological network and its complexity. In order to understand the high level meaning of a protein and a RNA function, we present the informational life of a protein and an RNA by zooming in their role in storing, carrying and interpreting information in different phases of their life. It would be easier to understand the complexity of the whole protein/RNA system, once we find an isomorphism between their information roles and a software system.

Protein and RNA are ubiquitously found in all the species we found on the earth without a single exception. All living organisms complete their life cycles through a series of biochemical reactions with the help of various proteins or RNAs. In the advanced life form, eukaryotic cells, the first part of a life cycle takes place in the nucleus. DNA, the genetic information carrier, duplicates itself. The information in DNA can be copied to RNA, when it needs to be passed outside nucleus. The information in RNA then is consumed in the ribosome for protein synthesis. That's a simplified version of central dogma (Crick, Francis. "Central dogma of molecular biology." Nature 227.5258 (1970): 561-563).

Beyond the central dogma, a real biochemical system consists of a large number of macromolecules and their interactions. These interactions include many interference and regulation on the way the information transferring from DNA to RNA and from RNA to protein synthesis. As a result, the amount of a protein in a living cell can increase or decrease dramatically within a short time due to a feedback loop. In some species of virus, the RNA can play the same role of DNA as a genetic information carrier. Thus the chain of biochemical reactions is shorter.

Complex Relationship

All these feedback and forward loops formed by biochemical reactions make the biological system the most complex system in human knowledge. In the past centuries of the scientific history, people invest very much to under this complex system, which will benefit the human health. The root cause of such complexity is the self-referential relationship or feedback loops, as we saw in many other complex systems.

We can compare the biological system with another complex system, software system, which is totally man-made and widely studied.

In the computer software world, data is a static object, which can be stored on hard disk or in computer memory. Program is dynamic, which has the capability to process data. The complexity of program is not because they can change data on hard disk or in memory, but change the behavior of themselves. It is the same way of proteins and RNAs regulating their production.

Data can be stored in computer in a serialized format or hierarchical structure. Programs are usually written in a specific syntax, which also implies their complexity. The world of data is convex, which means convex combination of two scalars is still a scalar. Two syntax correct C++ programs cannot be convex combined into another syntax correct C++ program. We can enumerate a 4x4 matrix with binary items, we can sample a random 8x8 matrix from a multivariate Gaussian distribution, but it is impossible to sample a random C++ program with correct syntax. It is also true for protein and RNA structures that we cannot enumerate all the possible structures of a protein or an RNA.

Protein/RNA structure is more close to program than data in a living cell. Crystalized protein/RNA has a static structure. They are the coordinates of all the atoms in a protein, i.e. (x,y,z) for each atom. In a living cell environment, a protein has the functionality to turn on or turn off some biochemical reactions. Such a switching mechanism makes them behave like the branching in a program. Protein structure defines which biochemical reaction a protein can change. Thus, the structure of protein molecules is program.

These observations are only valid on most high level programming languages. If we consult a high level language, LISP, it is not easy to draw the clear line between data and program. In Lisp language, the only syntax need to follow is that you need write everything as lists. A valid lisp program is a set of recursively-defined lists. This is just how RNAs play their roles in a living organism. It is both data and functional unit consuming genetic information stored in it.

To fulfill those complex roles, RNA and protein macromolecular evolves a great complexity in their structures. From strand to helices and sheets, their structures consist of many levels of sub-modules, like program libraries. By combining these sub-modules, protein and RNA molecular work together as a whole system. In the following parts, we will try to understand this complexity in RNA and protein structure by using a machine learning method.

## From Sequence to 3D Structure

From knowing that X-rays could be used to determine the structure of matter, 45 years passed before Max Perutz and John Kendrew solved the first protein 3D structure of a myoglobin protein, an iron and oxygen binding protein in the muscle tissue. This work gave the two authors the Chemistry Nobel Prize in 1962. Since then, the Chemistry Prize of 1964 was awarded for the structures of vitamin B12 and insulin, and the Chemistry Prize of 1988 was awarded for the structure of a membrane protein. Recently, milestone works include determining the structure of ribosome (Venki Ramakrishnan, Thomas A. Steitz, Ada Yonath, Chemistry Prize of 2009), a large complex of protein and RNA, and GPCR (G-protein-coupled receptors) (Brian Kobilka and Robert Lefkowitz, Chemistry Prize of 2012). In addition to X-ray crystallography, other laboratory methods include Nuclear Magnetic Resonance spectroscopy (NMR) and cryo-electron microscopy (EM), each of which works under different laboratory conditions.

As with many successes on protein and RNA structures determination using different laboratory methods, the technology on high-throughput sequencing advances also rapidly brings many more sequences with unknown structures. The speed of determining atom-level accurate structures by biological experiment methods always falls behind the speed of new sequence increases. For this reason, computational methods became popular for providing a good approximation solution to this problem. Molecular dynamic methods were invented more than 50 years ago (Giddings and Byring, 1955), but were not designed for macromolecular structures. Even with huge computational resources, people still cannot fold a whole protein molecule using the molecular dynamics method. In this thesis, we are focused on the homologous-based methods pioneered by (Bowie, et al., 1991; Rost, 1996; Subramaniam, et al., 1996; Sutcliffe, et al., 1987), which make use of known structures.

Functionality

There is a close relationship between the 3D structure and the functionality of a macromolecule, such as a protein or a RNA. The function of a protein includes its binding capability to other molecules, which can be determined by the shape of pockets on the surface of the protein. With a predicted structure, we can model the binding coefficient between a drug and its target, which helps evaluate the effect of the drug. 3D structures models can be used to estimate the sensitivity of a protein structure with regard to mutations in the sequence, which explains drug sensitivity differences for different people. 3D structures can also explain the stability of a given protein or RNA. There are several ways to assess the stability of a protein, which is needed to evaluate an artificial design of a protein sequence. More other applications include drug re-positioning, personalized diagnosis and genomic analysis.

Computational Challenge

The biggest challenge for protein and RNA 3D structure determination is the gap between the sequence data and the structure data. Though we have solved more structures, we have many more sequences without known structures still to be predicted. Another problem is how to determine and use appropriate methods to produce highly accurate results. A few experimental methods, such as the electron microscope, are available for discovering low accuracy 3D structures. However, it is not obvious how to integrate these results with current algorithms to improve 3D structure prediction accuracy.

All 3D structure prediction algorithms face much the same difficulty in that the searching space appears to be extremely large. To enumerate a structure, an algorithm should try all three possible coordinates for each atom in the space. In the coordinate representation, even if we consider only one atom for each amino acid or nucleotide, we have a search space of $R^3L$, where $L$ is length-which is the number of amino acids in a protein, or the number of nucleotides in a RNA. The coordinate representation of a macromolecular structure is too flexible without considering many macromolecule properties. This means that when we enumerate a structure from its coordinate representation, we will get an invalid or unreal structure in a large chance.

Structure search methods would be much more efficient, if we knew how to distinguish the physically valid decoys without biological meaning from those valid decoys. Due to the lack of complete rules for how a protein and RNA fold, we do not have efficient methods to test if a decoy is biologically meaningful by computing the energy function on the whole decoy. Thus, many decoys produced by the decoy sampling methods are rejected due to their poor quality.

In order to solve this complex problems, we decompose this big problem into several sub-problems, and develop various methods for them according to their own natures.

### *RNA 3D Structure*

RNA 3D structure has become an important research subject in recent years, and there is an increasing study of non-coding RNA in biology and health. Its growing important role appears in various life domains and processes, including regulating gene expression (Backofen *et al.*, 2007; Hiller *et al.*, 2007; Ray *et al.*, 2009; Solnick, 1985), interaction with other ligands (Badorrek *et al.*, 2006; Buck *et al.*, 2005) and stabilizing itself (Reymond *et al*., 2009). To elucidate the function of an RNA molecule, it is essential to determine its 3D structure. However, there are a great number of RNA sequences without solved structures. Experimental methods for RNA 3D structure determination are time-consuming, expensive and sometimes technically challenging. By far, there are ~29 million RNA molecules with (predicted) secondary structure in the Rfam database (Gardner *et al*., 2009), but only 4816 of them have tertiary structures in the nucleotide database (Berman *et al*., 1992). Therefore, we have to fill this large gap by predicting the 3D structure of an RNA using computational methods.

RNA tertiary structure prediction does not gain as much attention as secondary structure prediction (Akutsu, 2000; Alkan *et al.*, 2006; Backofen *et al.*, 2009; Bindewald and Shapiro, 2006; Eddy and Durbin, 1994; Ferretti and Sankoff, 1989; Gardner and Giegerich, 2004; Hamada *et al.*, 2009; Havgaard *et al.*, 2007; Hofacker, 2003; Knudsen and Hein, 2003; Mathews, 2006; Mathews and Turner, 2002; Mathews and Turner, 2006; Poolsap *et al.*, 2009; Will *et al.*, 2007; Zhang *et al.*, 2008; Zuker, 2003; Zuker and Sankoff, 1984). Both molecular dynamic methods (Bindewald and Shapiro, 2006; Hajdin *et al.*, 2010; Sharma *et al.*, 2008) and knowledge-based statistical methods (Das and Baker, 2007; Das *et al.*, 2010; Frellsen *et al.*, 2009) have been proposed to fold RNA molecules. The knowledge-based statistical methods for RNA tertiary structure prediction consist of two major components: an algorithm for conformation sampling and an energy function for differentiating the native structure from decoys. Fragment assembly, a knowledge-based method widely used for protein structure prediction (Haspel *et al.*, 2003; Lee *et al.*, 2004; Simons *et al.*, 1997), has been implemented in FARNA (Das and Baker, 2007) for RNA 3D structure prediction. However, this method has a couple of limitations: (i) there is no guarantee that any region of an RNA structure can be accurately covered by structure fragments in the RNA solved structure database, which currently contains only a limited number of non-redundant solved RNA structures; and (ii) sequence information is not employed in FARNA for conformation sampling. MC-Sym (Parisien and Major, 2008) is a motif assembly method for RNA 3D structure prediction, which uses a library of nucleotides cyclic motifs (NCM) to construct an RNA structure. MC-Sym has a time complexity exponential with respect to RNA length (i.e. the number of nucleotides), so MC-Sym may not be used to predict the tertiary structure for a very large RNA. As reported (Laing and Schlick, 2010), MC-Sym also fails in the case when the secondary structure of RNA lacks cyclic motifs. Recently, Frellsen *et al.* (Frellsen *et al*, 2009) have proposed a probabilistic model (BARNACLE) of RNA conformation space. BARNACLE uses a dynamic Bayesian network (DBN) to model RNA structures, but this DBN method does not take into consideration any sequence information. In addition, BARNACLE models the interdependency between the local conformations of only two adjacent nucleotides, but not of more nucleotides. Other RNA three dimensional structure prediction methods can be found in (Abraham *et al,* 2008; Das and Baker, 2007; Das *et al,* 2010; Ding *et al,* 2008; Flores *et al,* 2010; Frellsen *et al,* 2009; Gillespie *et al,* 2009; Hajdin *et al,* 2010; Jonikas *et al,* 2009; Laing and Schlick, 2010; Parisien and Major, 2008; Sharma *et al,* 2008; Tang *et al,* 2005; Wexler *et al,* 2006).

This study presents a novel probabilistic method conditional random fields (CRFs) (Lafferty *et al.*, 2001) to model RNA sequence–structure relationship. Different from BARNACLE modeling only RNA structures, our CRF method models the sophisticated relationship among primary sequence, secondary structure and 3D structure, which enables us to more accurately estimate the probability of RNA conformations from its primary sequence and thus sample RNA conformations more efficiently.

We have already successfully applied CRF to model protein sequence–structure relationship and conformation sampling (Zhao *et al.*, 2008, 2009, 2010). However, our CRF method for proteins cannot be directly applied to RNA. In order to apply CRF to RNA modeling, we have to employ a different method to represent an RNA 3D structure and model RNA bond torsion angles. We also have to face the challenge that there are a lot fewer solved RNA structures than the solved protein structures for CRF model training. By exploiting the secondary structure information of an RNA molecule, we have also developed a novel tree-based sampling scheme that can simultaneously sample conformations for two segments far away from each other along the RNA sequence. In contrast, our protein conformation sampling method can sample conformations for only one short segment at a given time. Finally, we also have to employ a totally different energy function for RNA folding. To the best of our knowledge, CRF has also been applied to RNA secondary structure prediction (Do *et al.*, 2006) and alignment (Sato and Sakakibara, 2005), but not modeling the relationship between RNA sequence and 3D structure.

Our method TreeFolder is more effective in sampling native like decoys than FARNA and BARNACLE, although we use the same simple energy function as BARNACLE, which contains only base-pairing information. Tested on 11 RNA molecules, TreeFolder obtains much better decoys for most of them. Our results imply that TreeFolder models RNA sequence–structure relationship well, which it is feasible to sample RNA conformations without using fragments and that sequence information is important for RNA conformation sampling. Experiments also show that TreeFolder works well with predicted secondary structures generated by tools such as CONTRAfold (Do *et al.*, 2006).

***Protein Secondary Structure***

Protein secondary structure (SS) is the local structure of a protein segment formed by hydrogen bonds and of great importance for studying a protein. There are two regular SS types: a-helix and b-strand, as suggested by Linus Pauling and his co-workers (Pauling et al, 1951) more than 50 years ago. A protein segment connecting a-helices and/or b-strands is called coil or loop region. Instead of using only Ca atoms Kabsch and Sander (Kabsch and Sander, 1983) group SSs into eight classes with all atom coordinates. This classification is a finer-grain model of the 3-class one and contains more useful information, such as the difference between 3-helix and 4-helix. The SS content of a protein is highly correlated with its tertiary structure and function. It is also suggested that SSs play the role of basic subunits during the folding and unfolding processes of a protein (Karplus and Weaver, 1994). Thus, SS can be considered as the transition state from primary sequence to tertiary structure. It will help in protein tertiary structure prediction and functional annotation if the SS is known (Myer and Oas, 2001). Protein SS prediction from primary sequence has been an important research topic in the field of bioinformatics. It is important to predict the 8-class SS since 8-class SS provides more detailed information about the local structure of a protein. Compared with 3-class SS, 8-class can tell 3-helix and 4-helix apart and describe different types of loop regions. This detailed description of protein SS is used in solving various protein structure problems (DeBartolo et al, 2009; Boden et al, 2006; Cheng and Baldi, 2007). In the case of identifying protein conformation changing (Boden and Bailey, 2006), using 8-class SS results in a better receiver operating characteristics curve than using 3-class. It is also much more challenging to predict the 8-class SS using machine learning methods because of the extremely unbalanced distribution of the 8-class SS types in native structures. A variety of machine learning methods have been proposed to predict protein SS (Pirovano and Heringa, 2010), especially for the 3-class (a, b or coil) SS prediction. For example, many neural Abbreviation: CNFs, conditional neural fields; CRF, conditional random field; HMM, hidden Markov model; NN, neural network; NR, non-redundant; PSSM, position-specific score matrix; SOV, segment overlap score; SS, secondary structure; SVM, support vector machines, neural network (NN) methods have been published for 3-class SS prediction (Qian and Sejnowski, 1988; Holley and Karplus, 1989; Kneller and Cohen, 1990; Rost and Sander, 1993; Rost and Sander, 1994; Rost 1996; Cuff and Barton 1999); these methods achieve Q3 accuracy of approximately 80%. PSIPRED is one of the representatives and is widely used for protein sequence analysis. However, these NN methods usually do not take the interdependency relationship among SS types of adjacent residues into consideration. Hidden Markov model (HMM) (Rabiner, 1989) is capable of describing this relationship (Asai et al, 1993; Aydin et al 2006), but it is challenging for HMM to model the complex nonlinear relationship between input protein features and SS, especially when a large amount of heterogeneous protein features are available. Support vector machines (SVM) have also been applied for SS prediction (Hua and Sun, 2001; Kim and Park, 2003; Ward et al, 2003; Guo et al, 2004; Duan et al, 2008). Similar to the NN methods, it is also challenging for SVM to deal with the dependency among SS types of adjacent residues. In addition, SVM outputs cannot be directly interpreted as or easily transformed into likelihood/probability, which makes the prediction results difficult to interpret. Very few methods are developed for 8-class SS prediction. To the best of our knowledge, SSpro8 (Pollastri 2002) is the only program that can predict 8-class SS of a protein. Similar to many other NN methods, SSpro8 does not exploit the interdependency relationship among SS types of adjacent residues.

In this research, we present conditional neural fields (CNFs) (Peng et al, 2009) method for 8-class protein SS prediction. CNF is a recently invented probabilistic graphical model and has been used for protein conformation sampling (Zhao et al 2009; Lafferty et al 2001) and handwriting recognition (Peng et al 2009). CNF is a perfect integration of CRF (conditional random fields) (Lafferty et al, 2001) and NNs and bears the strength of both CRFs and NNs. NNs can model the nonlinear relationship between observed protein features (e.g. sequence profiles) and SS types, but cannot easily model the interdependency among adjacent SSs. Similar to HMM, CRFs can model the interdependency among adjacent SSs, but cannot easily model the nonlinear relationship between observed protein features (e.g. sequence profiles) and SS types. By combing the strength of both NNs and CRFs, CNFs not only can model the nonlinear relationship between observed protein features and SS types, but also can model the interdependency among adjacent SSs. Similar to CRFs, CNFs can also provide a probability distribution over all the possible SS types of a given protein. That is, instead of generating a single SS type at each residue, our CNF method will generate the probability distribution of the eight SS types. The probability distribution may be useful for other purposes such as protein conformation sampling (Zhao et al, 2009; Zhao et al, 2008). Our CNF method achieves a much better Q8 accuracy than SSPro8.

### *Protein Contact Map*

Protein contact is defined as two residues of a protein are in contact if their Euclidean distance is less than 8 angstrom. The contact map is another way to represent a protein 3D structure, so it become a more and more important study topic. The distance of two residues can be calculated using $C_{alpha}$ or $C_{beta}$ atoms, corresponding to $C_{alpha}$ or $C_{beta}$ based contacts. A protein contact map is a binary L by L matrix, where L is the protein length. In this matrix, an element with value 1 indicates the corresponding two residues are in contact; otherwise, they are not in contact. Protein contact map describes the pairwise spatial and functional relationship of residues in a protein. Predicting contact map using sequence information has been an active research topic in recent years partially because contact map is helpful for protein 3D structure prediction (Ortiz et al., 1999; Vassura et al., 2008; Vendruscolo et al., 1997; Wu et al., 2011) and protein model quality assessment (Zhou and Skolnick, 2007). Protein contact map has also been used to study protein structure alignment (Caprara et al., 2004; Wang et al., 2013; Xu et al., 2007).

Many machine-learning methods have been developed for protein contact prediction in the past decades (Fariselli and Casadio, 1999; Gobel et al., 2004; Olmea and Valencia, 1997; Punta and Rost, 2005; Vendruscolo and Domany, 1998; Vullo et al., 2006). For example, SVMSEQ (Wu and Zhang, 2008) uses support vector machines for contact prediction; NNcon (Tegge et al., 2009) uses a recursive neural network; SVMcon (Cheng and Baldi, 2007) also uses support vector machines plus features derived from sequence homologs; Distill (Bau et al., 2006) uses a 2D recursive neural network. Recently, CMAPpro (Di Lena et al., 2012) uses a multi-layer neural network. Although different, these methods are common in that they predict the contact map matrix element-by-element, ignoring the correlation among contacts and also physical feasibility of the whole-contact map (physical constraints are not totally independent of contact correlation). A special type of physical constraint is that a contact map matrix must be sparse, i.e. the number of contacts in a protein is only linear in its length.

Two recent methods, PSICOV (Jones et al., 2012) and Evfold (Morcos et al., 2011), predict contacts by using only mutual information (MI) derived from sequence homologs and enforcing the aforementioned sparsity constraint. However, both of them demand for a large number (at least several hundreds) of sequence homologs for the protein under prediction. This makes the predicted contacts not useful in protein modeling, as a (globular) protein with many sequence homologs usually has similar templates in PDB; thus, template-based models are of good quality and hard to be further improved using predicted contacts. Conversely, a protein without close templates in PDB, which may require contact prediction, usually has few sequence homologs even if millions of protein sequences are now available. Further, these two methods enforce only a simple sparsity constraint (i.e. the total number of contacts in a protein is small), ignoring many more concrete constraints. To name a few, one residue can have only a small number of contacts, depending on its secondary structure and neighboring residues. The number of contacts of two strands is bounded by the strand length.

Astro-Fold (Klepeis and Floudas, 2003) possibly is the first method that applies physical constraints, which implicitly imply the sparse constraint used by PSICOV and Evfold, to contact map prediction. However, some of the physical constraints are too restrictive and possibly unrealistic. For example, it requires that a residue in one strand can only be in contact with a residue in another strand. More importantly, Astro-Fold does not take into consideration evolutionary information; thus, it significantly reduces its prediction accuracy.

In this study, we present a novel method PhyCMAP for contact map prediction by integrating both evolutionary and physical constraints using machine learning [i.e. Random Forests (RF)] and integer linear programming (ILP). PhyCMAP first predicts the probability of any two residues forming a contact using evolutionary information (including MI), predicted secondary structure and distance-dependent statistical potential. PhyCMAP then infers a given number of top contacts based on predicted contact probabilities by enforcing a set of realistic physical constraints on the contact map. These restraints specify more concrete relationship among contacts and also imply the sparsity restraint used by PSICOV and Evfold. By combining both evolutionary and physical constraints, our method greatly reduces the solution space of contact map and leads to much better prediction accuracy. Experimental results confirm that PhyCMAP outperforms currently popular methods no matter how many sequence homologs are available for the protein under prediction.

The following parts of this thesis are organized as following. Chapter 1 introduces background, motivation, and challenges of related studies. Chapter 2 lists basic concepts and problem definitions. Chapter 3 is about how to predict intermediate states, in both protein and RNA. Chapter 4 presents our result applying intermediate prediction to 3D structure prediction. Chapter 5 concludes with the difficulty and challenge of this study.

# Chapter 2. Problem Definition and Basic Concepts

## Formularization of 3D structure

For a protein or a RNA molecule, its 3D structure consists of all the coordinates of all its atoms. We consider a protein of a few hundred amino acids as an example. For simplicity, we can use the coordinates of the alpha carbon in each amino acid as a delegate of the position of the amino acid. This assumption is from the observation that each amino acid has a relatively fixed 3D structure. Thus, the 3D structure of a protein is described by all the coordinates of the alpha carbons, or backbone atoms. For a RNA case, the difference is that each nucleotide has six atoms, so we use two atoms for each nucleotide as its backbone atoms in order to keep the representation simple and accurate (Cao, 2005; Jonikas 2009).

Using the coordinates to describe 3D structures is good for visualization, however, is not a good idea for predictive modeling. This is for two reasons. First, the coordinates change with the coordinating systems we choose. This introduces difficulty when building comparable samples from different protein structures used in training a machine learning model. Secondly, the coordinates of amino acids, which are not independent variables, are quite tangled with each other. The distance of two neighboring amino acids has a nearly fixed value, which is a strict constraint between coordinates. If we built a learning model to directly predict the coordinates, we would have to consider all the constraints in the prediction step. For a 3D structure, the coordinates of each atom make not much sense without considering its relative position to other atoms.

A better representation of the 3D structure is the rotating dihedral angles along the chain of a protein or a RNA. The dihedral angle for a position on a sequence is computed between the two neighbor planes formed by each of four backbone atoms. The dihedral angles are invariant with the coordinating system we choose, and keep all the relative positions between bone atoms. Another approach to represent the 3D structure is a fragment library method. This method includes a fragment library, which is an assembly of real 3D structure fragments of very short sequences. This method builds a whole 3D structure decoy by collaging many fragments together. Thus, it excludes any structures with an unphysical fragment, however, many valid structure decoys have also been excluded from the representation, especially those structures with an unknown structure fragment.

Intermediate states

The folding process from the beginning of the sequence to the final functional structure includes several phases and transient states, which is still unclear in detail for most proteins and RNA. Generally, the folding process spreads from neighbored amino acids or nucleotides groups to the whole sequence. From the functional structure determined by experiments, scientists also observed patterns of local 3D structures. These patterns may be associated with the intermediate states of protein folding, which include protein and RNA secondary structures and the protein contact map. Knowing these intermediate states for each local amino acid or nucleotides group does not suggest the accurate 3D structure of all atoms. However, intermediates states, such as a secondary structure or a contact map, are much simpler than the 3D structure and usually possess a nicer mathematical form. This simplicity invites a wider range of researchers to tackle the problem of predicting the secondary structure over that of 3D structures. In turn, the results of intermediate structure prediction can help improve accuracy of 3D structure prediction.

Features

Before describing the detailed method we proposed for protein and RNA structure modeling, we want to present the definitions of some common symbols and notations used in the following sections.

*Italic:* Non-text notations, which can be variables, e.g. *L* ; or atom name, e.g. $C_\alpha$, or secondary structure types

Å: Angstrom

$C_\alpha$: Alpha carbon in an amino acid

$C_\beta$: Beta carbon in an amino acid

*H, E, C*: α-helix, β-strand, and coil as secondary structure types

Sequence information:

In the following paragraphs, we denote *L* as the length of the query sequence.

PSSM, position specific scoring matrix, is the mutation probability for each amino acid or nucleotide in the sequence of a protein or a RNA. For each position in the sequence, the mutation probability is a row vector indicating the probability for this position to mutate to other types of amino acids or nucleotides. The number of the rows in a PSSM is the length of the protein or RNA. Thus, for a sequence with the length of *L*, the matrix will be *L* x 20 or *L* x 4 for protein and nucleotide, respectively. This matrix can be produced by using multiple sequence alignment programs. In our research, we use PSI-BLAST (Schäffer, et al., 2001) to compute the PSSM for protein sequences.

Distance Matrix: an *L* by *L* matrix, each element at *(i,j)* is the distance of the atom *i* and the atom *j* in angstrom.

Contact: An atom pair has a contact if their distance is less than 8Å. The contact between two amino acids is defined as the contact between their representative atoms. There are 3 ways to choose this representative atom of the amino acid, $C_\alpha$, $C_\beta$, or the closest two atoms of a pair of amino acids. An alternative threshold of the distance is 6Å, which is not commonly used.

Contact Map: an $L$ by $L$ binary matrix to describe if there is a contact for each atom pair in the sequence. This binary contact can be relaxed into a probability. In the probability matrix, each element is a [0,1] real value to indicate the probability of whether the two atoms have a contact.

Secondary Structure of a Protein: The secondary structure of a protein is a sequence of secondary structure elements, which can be 3-class or 8-class. Each element in the secondary structure sequence describes the local sub-structure pattern around the amino acid at the corresponding position in the protein sequence.

Secondary Structure of RNA: The secondary structure of RNA describes the contact map of RNA. It is a binary matrix, where each element represents whether two nucleotides are close in space and have a hydrogen bond between them.

# Chapter 3. Intermediate states bridging sequences and 3D structures

The whole life of the folding process of a protein/RNA molecule can be sliced into many different phases. Short range interactions formed after the sequence is assembled. There are then also long range interactions. Those interactions can be grouped into several patterns, which are the intermediate states, including protein/RNA secondary structures and protein contact maps. On the side of protein, short-range interactions form secondary structures, and long range interactions are associated with a protein contact map. On the side of RNA, the RNA secondary structure means there are both short range and long range interaction patterns.

## 3.1 Protein Secondary Structure

Compared with the protein 3-class secondary structure (SS) prediction, the 8-class prediction gains less attention and is also much more challenging, especially for proteins with few sequence homologs. This paper presents a new probabilistic method for 8-class SS prediction using conditional neural fields (CNFs), a recently invented probabilistic graphical model. This CNF method not only models the complex relationship between sequence features and SS, but also exploits the interdependency among SS types of adjacent residues. In addition to sequence profiles, our method also makes use of non-evolutionary information for SS prediction. Tested on the CB513 and RS126 data sets, our method achieves Q8 accuracy of 64.9 and 64.7%, respectively, which are much better than the SSpro8 web server (51.0 and 48.0%, respectively). Our method can also be used to predict other structure properties (e.g. solvent accessibility) of a protein or the SS of RNA.

Each protein secondary structure represents a simplified local pattern of a protein 3D structure. The local pattern involves amino acids that are consecutive within the sequence. To present a simplified classification, there are 3 secondary structure types, α-helix, β-strand, and coil, as suggested by Linus Pauling and his co-workers more than 50 years ago (Pauling, *et al.*, 1951). Among the three secondary structure types, helices and strands have a pattern, but coils are defined as the unclassified parts between two helices and strands without a stable pattern. This 3-class model is extended by Kabsch and Sander group to 8 classes by adding sub-classes for each secondary structure class in the 3-class system (Kabsch and Sander, 1983). By using an 8-class classification, we have more information to distinguish the differences of 4-helix and 3-helix, and the differences among coil patterns.

We consider protein secondary structure as a kind of intermediate state due to both practical reasons and biological evidence. The secondary structure of a protein provides partial information about its 3D structure other than the sequence. Thus, it is used by biologists to identify protein function (Myers and Oas, 2001). It is also the basic unit in the process of the folding and unfolding of a protein (Karplus and Weaver, 1994), where the secondary structures are an intermediate phase between the state of a sequence and the state of a 3D structure.

There is research predicting protein secondary structures based on the protein sequence information, including the sequence and PSSM (Pirovano and Heringa, 2010). Most methods focus on 3-class prediction. Neural network methods (Cuff and Barton, 1999; Holley and Karplus, 1989; Jones, 1999; Kneller, et al., 1990; Qian and Sejnowski, 1988; ROST, 1996; Rost and Sander, 1993; Rost and Sander, 1994) achieve Q3 accuracy of ~80%, among which PSIPRED is the most representative one. These methods take input features from each independent position of the sequence, and do not model the transition relationship from one position the next. The Hidden Markov Model can capture this transition, but cannot model the non-linear relationship between sequence features and secondary structure labels. To fill this gap, we model this problem by a conditional neural field model, which is a probabilistic graphical model taking advantage of neural networks and conditional random fields.

Compared with predicting 3-class secondary structures, the 8-class prediction problem is more challenging and more important. The 8-class secondary structure of a protein will provide more information than the 3-class one. It distinguishes the difference between 3-helix and 4-helix and recognizes different types of loop regions. To predict an 8-class structure is much more difficult than predicting a 3-class one, because the distribution of 8-class secondary structures is extremely unbalanced in the training dataset.

In the results of applying our 2$^{nd}$ order conditional neural fields on the 8-class prediction problem, we have significantly improved the 8-class secondary structure prediction. Our method of second order conditional neural fields can be illustrated by Figure 3-1-1.

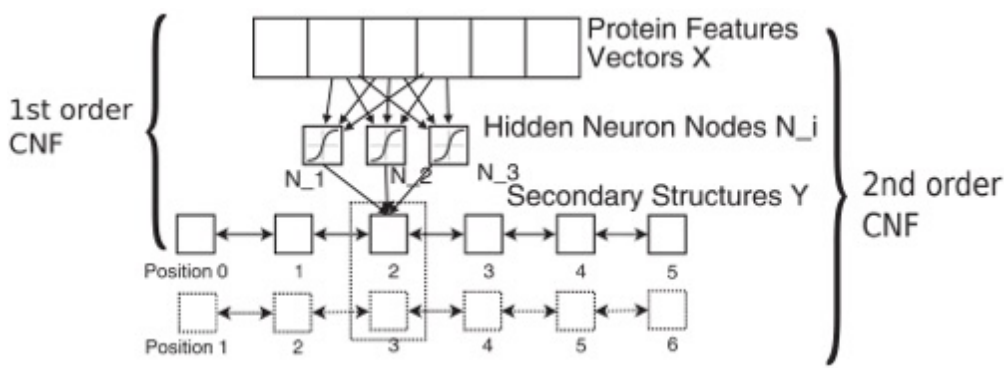

**Figure 3-1-1. Conditional neural field model for an 8-class secondary structure prediction. The sigmoid neural network layer transforms the input feature in a non-linear way, and the two order conditional random field models neighborhood dependency as well as the output of the neural network.**

### *Method*

We represent the input features of a given protein by an $N \times L$ matrix $X = (X_1, \ldots X_L)$, where $L$ is the number of residues in the protein. The $k$th column vector $X_k$ represents the protein features associated with the $k$th residue. We represent the secondary structure of a protein using a sequence of integers taking values from one to the number of secondary structure types (i.e. 8). Formally, for a given protein with length L, we denote its secondary structure as $Y = (Y_1, \ldots, Y_L)$, where $Y_i \{1, 2 \ldots, 8\}$.

In Equation (3-1-1), our model defines the conditional probability of SS Y on protein features X as follows:

$$P(Y|X) \propto \exp\left[\sum_{i=1}^{L-1} \psi(Y_i, Y_{i+1}) + \sum_{i=1}^{L} \sum_{j=1}^{m} \phi(Y_i, N_j\left(X_{i-\frac{k}{2}}, \ldots, X_{i+\frac{k}{2}}\right))\right] \quad (3\text{-}1\text{-}1)$$

Here, $N_j()$ is a hidden neuron function that does nonlinear transformation of input protein features, k is the window size, and m is the number of hidden neuron nodes (i.e. $N_j()$). The edge feature function $\varphi()$ models the interdependency between two adjacent SS types and the label feature function $\varphi()$ models the dependency of SS type on a window (with the size of *k*) of sequence information. Formally, $\psi()$ and $\varphi()$ are defined as follows:

$$\psi(Y_i, Y_j) = \sum_{a,b} t_{a,b} I(Y_i = a) I(Y_j = b)$$

$$\phi(Y_i, N_j) = w_{i,j} N_j + \sum_a u_a I(Y_i = a) \quad (3\text{-}1\text{-}2)$$

*I()* is an indicator function, *w, u,* and t are model parameters to be trained and a and b represent SS types. The formula in Equation (3-1-1) is explained in Figure 3-1-1, which contains three layers of nodes: the SS types $Y_i$, the hidden neurons $N_i$, and the input protein feature vectors *X*. The conditional probability *P(Y|X)* depends on X and the output values from the hidden neuron nodes. Neuron nodes extract information from a shifting window of *X*. To capture higher order dependency among adjacent SS types, we combine SSs of two adjacent residues into a single second-order type, which results in the second-order CNF model. Formally, we use $\tilde{Y}_i = (Y_i, Y_{i+1})$, *i = 1, ..., L – 1* as the second-order type on position *i* instead of $Y_i$. With this transformation, we obtain a second-order CNF model.

### *Model training and prediction*

***Training*** Here, we only briefly introduce the training algorithm for CNF. Please refer to (Peng et al, 2009) for a detailed information of the CNF training algorithm.

Given $K$ training proteins with features $U_k = (X_1^k, \ldots, X_{L_k}^k)$ and their native SS types $V_k = (Y_1^k, \ldots, Y_{L_k}^k)$, $k = 1, \ldots, K$, we train the CNF model by maximizing the occurring probability of the native SS types. That is, we estimate the parameters $w$ and $v$ in Equation (3-1-2), and by maximizing $f = \prod_{k=1}^{K} P(Y_k | X_k)$. Empirically, to avoid overfitting caused by a large number of model parameters, we add a regularization factor into the log-likelihood. That is, instead of maximizing $f$ we maximize $\log f + \gamma ||w, t, u||_2$, where $\gamma$ is a regularization factor used to control the model complexity. Although we cannot guarantee that the solution found is the optimal solution in this training problem, we can find a quite good suboptimal solution using the LBFGS (limited memory BFGS, (Liu and Nocedal 1989)) algorithm. To obtain a good final solution, we can generate a set of suboptimal solutions with different starting solutions and use the best suboptimal solution as the final solution. The best window size $k$ and regularization factor $\gamma$ can be determined using cross-validation.

**Prediction** After the model parameters are trained, we can use the model to predict the SS of a protein. We first use a forward-backward algorithm (Rabiner 1989) to calculate the marginal probability of eight SS types at each residue, $P(Y_i|X)$. Then the SS type with the highest marginal probability can be used as the predicted SS.

### Protein features

We use both position-dependent and position-independent protein features: PSSM (position-specific score matrix), propensity of being endpoints of SSs, physico-chemical property, correlated contact potential of amino acids, and primary sequence. The first one feature is position-dependent and the latter four are position-independent.

PSSM contains evolutionary information derived from sequence homologs and is a position-dependent feature. The other three features are position-independent and not directly relevant to evolutionary information. To produce PSSM of a given protein, we use PSIBLAST to search it against the NR (non-redundant) database with E-value = 0.001 and five iterations. Low information and coiled-coil sequences in the NR database are removed as outliers using the pfilt program in the PSIPRED package.

Every amino acid has a specific propensity of being endpoints of an SS segment. Duan et al (Duan et al, 2008) demonstrated that it helps in SS prediction by using this kind of propensity. We generated this kind of propensity by a simple statistics on a set of NR protein structures.

We also use the physico-chemical property of an amino acid as the input features. The physico-chemical property has been studied in (Meiler et al, 2001) for SS prediction and is represented by a vector of seven scalars.

The matrix of correlated contact potential of amino acids estimates the correlation between two amino acids by calculating the correlation of the contact potential vectors of these two amino acids. The contact potential of an amino acid is taken from (Tan et al, 2006). The correlated contact potential matrix has been used by Peng and Xu (Peng and Xu, 2010) for finding templates in protein tertiary structure prediction and has proved to be useful.

The information contained in a primary sequence can be represented by a 20 × 20 identity matrix, where each unit row vector represents an amino acid. Thus, the primary sequence is denoted as the identity matrix in the following sections.

In summary, every residue has 20-dimension PSSM features and 58-dimension position-independent features. In the experiments shown in Tables 3-1-1 and 3-1-2 and Section 3.3, we use all of those input features. The native SS of a given protein is calculated using the DSSP package (Kabsch 1982).

**Table 3-1-1. Q8 accuracy of our CNF method and SSPRO8 on the CB513 data set. Q-H, Q-G, Q-I, Q-E, Q-B, Q-T, Q-S, and Q-L represent**

the prediction accuracy on a single SS type H, G, I, E, B, T, S, and L respectively. Bold indicates improvement.

|  | CNF method | | SSpro8 | |
|---|---|---|---|---|
|  | Mean | Std | Mean | Std |
| Q8 | **0.633** | 0.013 | 0.511 | 0.015 |
| Q-H | **0.887** | 0.009 | 0.752 | 0.031 |
| Q-G | **0.049** | 0.015 | 0.007 | 0.002 |
| Q-I | 0 | 0 | 0 | 0 |
| Q-E | **0.776** | 0.016 | 0.597 | 0.013 |
| Q-B | 0 | 0 | 0 | 0 |
| Q-T | **0.418** | 0.008 | 0.327 | 0.017 |
| Q-S | **0.117** | 0.015 | 0.049 | 0.003 |
| Q-L | **0.608** | 0.014 | 0.499 | 0.019 |
| SOV | **0.206** | 0.025 | 0.141 | 0.015 |

Table 3-1-2. The average Q8 accuracy and the sensitivity of each subclass tested on CB513 and RS126 using five CNF models trained from the CULLPDB data.

|  | On all data of CB513 | | | On all data of RS126 | | |
|---|---|---|---|---|---|---|
|  | CNF method | | SSpro8 | CNF method | | SSpro8 |
|  | Mean | Std |  | Mean | Std |  |
| Q8 | **0.649** | 0.003 | 0.51 | **0.647** | 0.003 | 0.48 |
| Q-H | **0.875** | 0.004 | 0.752 | **0.9** | 0.005 | 0.728 |
| Q-G | **0.207** | 0.021 | 0.006 | **0.229** | 0.022 | 0.016 |
| Q-I | 0 | 0 | 0 | 0 | 0 | 0 |
| Q-E | **0.756** | 0.001 | 0.598 | **0.797** | 0.002 | 0.577 |
| Q-B | 0 | 0 | 0 | 0 | 0 | 0 |
| Q-T | **0.487** | 0.004 | 0.33 | **0.488** | 0.01 | 0.308 |
| Q-S | **0.202** | 0.009 | 0.051 | **0.153** | 0.009 | 0.051 |
| Q-L | **0.601** | 0.002 | 0.5 | **0.614** | 0.006 | 0.479 |
| SOV | **0.194** | 0.002 | 0.143 | **0.18** | 0.003 | 0.123 |

Given the PSSM of a protein, we calculate a value (Neff) for each residue in the protein to evaluate the information content derived from sequence homologs

$$N_{eff} = \exp\left(\sum_{i=1}^{20} -p_i \log p_i\right) \quad (3)$$

In the above formula, $p_i$ is the frequency vector converted from PSSM for the *i*th residue in the protein. The N$_{eff}$ value of a given protein is calculated by the average N$_{eff}$ of all the residues. The Neff of a protein approximately measures the number of NR sequence homologs that can be detected by PSIBLAST from the NR database. Neff ranges from 1 to 20 since there are only 20 amino acids in protein sequences. We call those proteins with small Neff values (e.g. <5) as low-homology proteins since they do not have a large number of NR sequence homologs in the NR database. According to our experience, the more NR sequence homologs are available for a protein, the easier to predict its SS. It is very challenging to predict SS for the proteins with few sequence homologs.

### *Training and test data sets*

We use two public benchmarks CB513 (Cuff, 1999) and RS126 (Rost 1993) to test the performance of our CNF method and study the relative importance of various features.

We also use a large data set cullpdb_pc30_res2.5_ R1.0_d100716 (denoted as CullPDB) from the PISCES server (Wang and Dunbrack, 2003), which contains about 8000 proteins with high resolution structure (<2.5Å) and up to 30% sequence identity. We also remove protein chains with less than 50 or more than 1500 residues. For those chains with missing residues, we cut them at the missing positions into several segments. The redundancy between the CullPDB set and CB513 and RS126 is also removed.

In our results, all cross-validation we conduct is fivefold cross-validation; 4/5 of the data is used for training and 1/5 of the data for validation. In the experiments of Table 3-1-1, and Figure 3-1-2, the results are averaged among fivefold cross-validation on their data sets. In Table 3-1-2 and Figure 3-1-3, CNF models are trained on the data set of CullPDB. We do not do cross-validation on the data set of RS126, since it only contains 126 proteins.

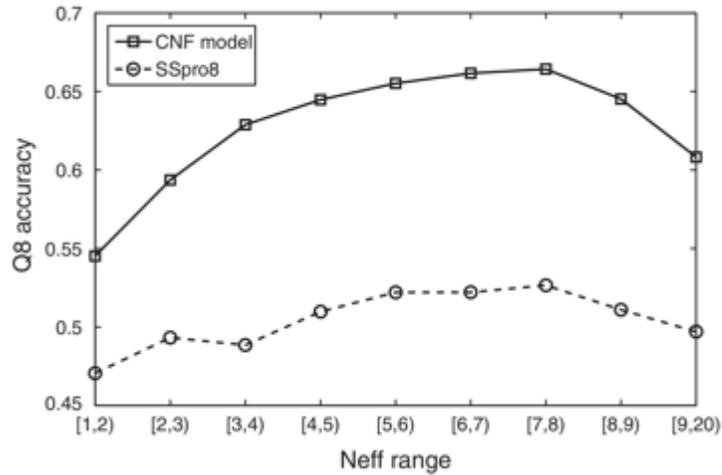

**Figure 3-1-2. Q8 accuracy between our CNF model and SSpro8 on the CB513 dataset. The accuracy is plotted with the Neff ranges of the sequences.**

## *Results*

### *Q8 accuracy and SOV on CB513 and RS126*

We use Q8 accuracy and SOV (segment OVerlap score) (Rost et al, 1994) to compare our CNF method with SSPro8 on two data sets, CB513 (Cuff, 1999) and RS126 (Rost 1993). To evaluate the Q8 accuracy of SSPro8, we submit the proteins sequences to the SSpro8 server and parse results returned by the server. To evaluate the Q8 accuracy of our method, we employed two strategies. One is to conduct cross-validation on CB513. The other is to train our CNF model using the CullPDB data set and then test the CNF model on CB513 and RS126. (Redundancy between the CB513/RS126 data set and the CullPDB set is removed.) Table 3-1-1 shows the overall Q8 accuracy of our CNF method and SSPro8 and their accuracy on each SS type as well as SOV (Rost et al, 1994). This table shows that our CNF method significantly outperforms the SSpro8 web server. Table 3-1-2 shows results on all data of CB513 and RS126 data sets. In this table, five CNF models are trained from CullPDB with different initializations and a consensus prediction is made for each residue. The accuracy on all data of CB513 is higher than the Q8 in Table 3-1-1, maybe because more data are used in training. On both CB513 and RS126, our method outperforms SSpro8 significantly.

We also conduct a cross-validation test on the CullPDB data set and achieve 67.9% Q8 accuracy in average. The confusion matrix shows most 3/10-Helices(G) are predicted as α-Helices(H), Turns(T), and Loops(L). β-Bridges(B) are likely to be predicted as Extended strands(E) or Loops(L). Turns(T) have a high propensity to be misclassified as α-Helices(H). Bends(S) are probably to be predicted as Loops(L) and Turns(T). There are also another two types of confusions. One is the confusion inside the same type of SSs, such as H, G are both helices, and E, B are both strands. The other is overlap between different classes. A Turn(T) is defined in [2] as one amino acid has a hydrogen bond with another, but not in a helix. This definition implies the similarity between a Turn and a Helix.

### *Relative importance of various features*

There is no doubt that the PSSM is the most important information for SS prediction. Here, we examine the relative importance of various position-independent features for SS prediction. Our analysis shows that the propensity of being SS endpoints contributes more than other features.

We compare those Q8 accuracy values of four position-independent features (physicochemical property, propensity of SS endpoints, correlated contact potentials, and identity matrix) without PSSM features. No matter what the regularization factor is, the feature of SS endpoints works better than other features.

However, when PSSM is also used, there seems to be no obvious difference among these position-independent features. Figure 3-1-2 shows Q8 accuracy of our CNF method with different features and trained with different regularization factors. Among the highest accuracy of all the CNF models, the gap is <1% in Q8 accuracy. In this figure, Test F-1 uses merely PSSM features, Test F-2 uses PSSM and Neff, Test F-3 uses PSSM and physicochemical features, Test F-4 uses PSSM and SS endpoints, Test F-5 uses PSSM and contacts potentials, and Test F-6 uses PSSM and identity matrix. Every accuracy value is the average of five repeated experiments with same model parameters. The Q8 accuracy values in Figure 3-1-2 of different features reach their maximum with different regularization factors, but are not far from each other, which indicates that position-independent features do not help improve 8-class SS prediction if PSSM is used.

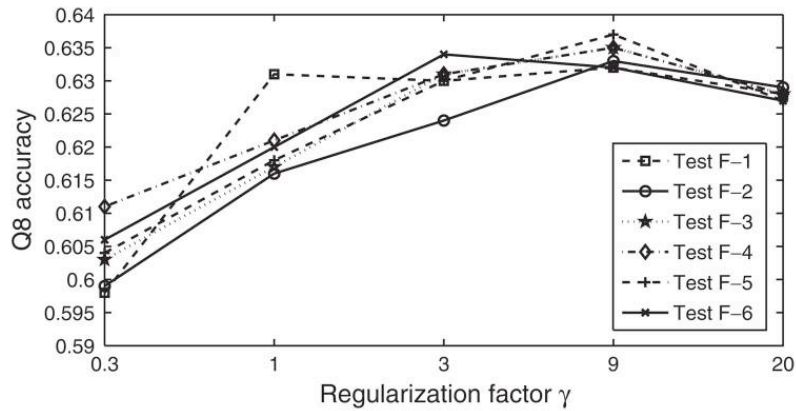

**Figure 3-1-3. Q8 accuracy of our CNF method on the data set of CB513 with different position-independent features combined with PSSM features and trained with different regularization factors.**

*Q8 accuracy on proteins with various Neff values*

We also examine the performance of our method with respect to the Neff value of a protein. As shown in Figure 3-1-3, we divide the Neff values into nine intervals and calculate Q8 accuracy of both our CNF method and SSPro8 on CB513 with respect to Neff. On average our method performs better than SSPro8 no matter which Neff interval is considered. Our CNF model performs best on the proteins with Neff values in [7,8). The performance of our CNF method increases with respect to Neff when it is <8. It is interesting to notice that the performance decreases when Neff is larger than 8. This may imply that for SS prediction, it may not be the best strategy to use evolutionary information in as many homologs as possible. Instead, we should use a subset of sequence homologs to build sequence profile when there are many sequence homologs available. That is, we should not use a sequence profile with Neff larger than eight to predict SS. One possible method to reduce Neff is to run PSIBLAST with smaller E-value or fewer iterations.

*Conclusions*

We have presented a method for 8-class SS prediction using CNFs. Our CNF model not only captures the complex nonlinear relationship between protein features and SS, but also exploits interdependence among SS types of adjacent residues. This is why we can achieve a much better performance than the existing methods. Furthermore, our CNF model defines a probability distribution over the SS space. The probability distribution provided by our method can be in turn applied to protein conformation sampling (Zhao 2008; Zhao, 2009).

The error in SS prediction may result from various factors. Similar to most existing methods, our method does not take into consideration long-range inter-residue interaction, which may be important for β-sheet prediction. This may be the major reason why we cannot do prediction on the β-strand as well as the α helix. The unbalanced distribution of SS types also makes it challenging to predict 8-class SS, especially for 3/10-helix, β-bridge and π-helix. These SS types in total account for only around 5% of residues.

Our method also achieves Q3 accuracy 81.17% on the CullPDB data set, comparable to the popular three-class SS prediction program PSIPRED. We can further improve our method by improving the wiring scheme used to connect input features to hidden neurons so as to extract more information from sequence profiles and the chemical property profile of amino acids. It may also help by increasing the number of output states in our CNF model. As discussed in this paper, amino acids have different propensities of being in the two ends of a single SS segment. Therefore, we can split an SS state into several subcategories, like the head of a helix, the tail of a helix, and the middle of a helix. We examine our method by comparing it with other state-of-the-art methods, including SSpro8 (Pollastri, et al., 2002). Our numerical results show the significant improvement of our method and explain how the Neff value affects performance of the secondary structure prediction.

## 3.2 Protein contact map

Another intermediate state in the process of the protein sequence folding into a protein 3D structure is the contact map. Compared with the local pattern in the secondary structure, the protein contact map is a representation of long-range patterns in a 3D structure.

The protein contact map can be written as a binary matrix, $M$, where $M_{i,j}=0$ means the $i$th amino acid and the $j$th amino acid have no contact, i.e. their distance is larger than 8Å. Also, $M_{i,j}=1$ if and only if they have a contact. A contact between two amino acids usually implies a functional relationship, therefore the prediction of the protein contact map is an important topic for structural biology (Ortiz, et al., 1999; Vassura, et al., 2008; Vendruscolo, et al., 1997; Wu, et al., 2011).

There are two main pitfalls in most machine learning methods have been tried to the protein contact prediction problem. The first is the failure of modeling dependency between amino acid pairs. This dependency is very common for two beta strands forming a beta sheet or for two helices forming a parallel structure. The second is the difficulty to model the sparseness of a contact matrix. A protein contact map is sparse and the total number of contacts is in a linear scale with the length of the sequence. Neither dependency nor sparseness had been considered by all machine learning methods by the time we invented own contact prediction method.

Other than a parametric model, the mutual information is found to be a method to predict protein contact map (Jones, et al., 2012; Morcos, et al., 2011), which measures the co-mutation relationship of each pair of amino acids. The problem with these methods is that they do not use template-based information. For most globular proteins, the template information can provide a very accurate prediction of 3D structures. In the other case, proteins without a good homologous structure it will be difficult to use these two methods, which are heavily dependent on the homologous search. These two methods also do not model the dependency between amino acid pairs.

The first method that systematically includes the amino acid pair-wise dependency, is Astro-Fold (Klepeis and Floudas, 2003), which model the problem as an integer programming problem and encodes the dependency in linear constraints. However, Astro-Fold does not make use of evolutionary information, such as PSSM, which has proved very useful in many protein structure studies.

In order to take advantage of the homologous information of a protein sequence as well as to model the dependency of amino acid pairs, we create a PhyCMAP, which is a mixture model integrating a Random Forest and integer linear programming. In the first step of PhyCMAP, the random forests model predicts the contact probability between each pair of amino acids. The probability is the input of the second step, the integer linear programming model, which output a filtered probability after removing the conflicts among amino acid pairs caused by the contact dependency.

Although studied extensively, it remains challenging to predict contact map using only sequence information. Most existing methods predict the contact map matrix element-by-element, ignoring correlation among contacts and physical feasibility of the whole-contact map. A couple of recent methods predict contact map by using mutual information, taking into consideration contact correlation and enforcing a sparsity restraint, but these methods demand for a very large number of sequence homologs for the protein under consideration and the resultant contact map may be still physically infeasible. This study presents a novel method PhyCMAP for contact map prediction, integrating both evolutionary and physical restraints by machine learning and integer linear programming. The evolutionary restraints are much more informative than mutual information, and the physical restraints specify more concrete relationship among contacts than the sparsity restraint. As such, our method greatly reduces the solution space of the contact map matrix and, thus, significantly improves prediction accuracy. Experimental results confirm that PhyCMAP outperforms currently popular methods no matter how many sequence homologs are available for the protein under consideration.

## *Method*

As shown in Figure 3-2-1, our method consists of several key components. First, we use RF to predict the contact probability of any two residues based on a few protein features related to these two residues. Then we use an ILP method to select a set of top contacts by maximizing their accumulative probabilities subject to a set of physical constraints. The resultant top contacts form a physically favorable contact map for the protein under consideration.

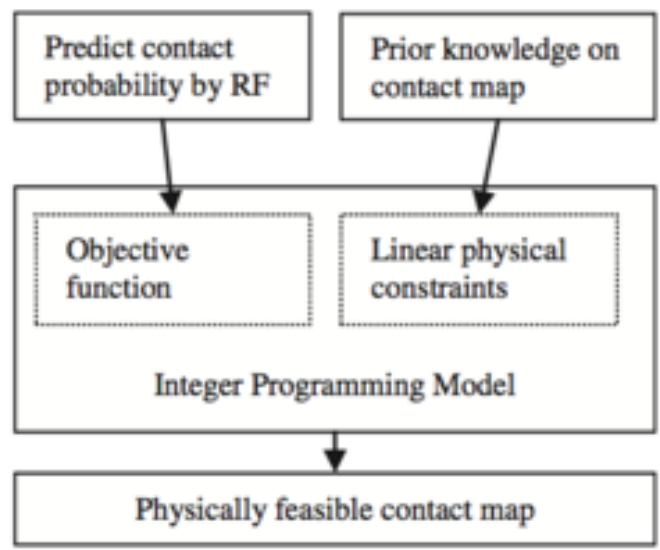

**Figure 3-2-1. The workflow of our protein contact map prediction method.**

## *Predicting contact probability by Random Forests*

We use RF to predict the probability of any two residues forming a contact using the following input features: EPAD (a context-specific distance-based statistical potential) (Zhao and Xu, 2012), PSIBLAST sequence profile (Altschul and Koonin, 1998), secondary structure predicted by PSIPRED (Jones, 1999), pairwise contact score and contrastive MI (CMI) derived from multiple sequence alignment (MSA) of the sequence homologs of the protein under prediction. The latter four features, including EPAD energy scores, sequence profile, pairwise contact score, and contrastive mutual information, are calculated on the residues in a local window of size 5 centered at the residues under consideration. In total, there are 300 features for each residue pair. We trained our RF model using the Random Forest package in R (Breiman, 2001; Liaw and Wiener, 2002) and selected the model parameters by 5-fold cross-validation.

EPAD energy score is the context-specific interaction potential of the C or C atoms of two residues at all the possible distance bins is used as features. The atomic distance is discretized into some bins by 1 angstrom, and all the pairs with distance larger than 15 angstrom are grouped into a single bin.

Sequence profile feature is the position-specific mutation scores at residues *i* and *j* and their neighboring residues are used. In additional to position specific scores, a protein contact-based potential CCPC (Tan et al., 2006) and amino acid physic-chemical properties are also used as features of our RF model.

Homologous pairwise contact score (HPS) is the feature calculated from two residues *i* and *j* of the protein under consideration. Let H denote the set of all the sequence homologs. Given an MSA of all the homologs in H, we calculate the homologous pairwise contact score HPS for two residues *i* and *j* as follows.

$$HPS(i,j) = \sum_{h \in H} PS_{beta}(a_i^h, a_j^h) + PS_{helix}(a_i^h, a_j^h)$$

In the definition above, $a_i^h (a_j^h)$ denote the residue in a homolog *h* aligned to residue *i(j)* in the query sequence. $PS_{beta}(a_i^h, a_j^h)$ is the probability of two amino acids $a^h_i$, $a^h_j$ forming a contact in a beta-sheet. $PS_{helix}(a_i^h, a_j^h)$ is the probability of two amino acids $a^h_i$, $a^h_j$ forming a contact connecting two helices. The probability is calculated as follows.

$$PS_{beta}(a_i^h, a_j^h) = \frac{|(A,B) \text{ forming a beta contact}|}{\text{Total number of beta contact pairs}}$$

$$PS_{helix}(a_i^h, a_j^h) = \frac{|(A,B) \text{ forming a helix contact}|}{\text{Total number of helix contact pairs}}$$

The contrastive mutual information feature describes the dependency between two residues mutation. Let $m_{i,j}$ denote the MI between these two residues $i$ and $j$, which can be calculated from the MSA of all the sequence homologs. We define the CMI as the MI difference between one residue pair and all of its neighboring pairs, which can be calculated as follows.

$$CMI_{i,j} = (m_{i,j} - m_{i-1,j})^2 + (m_{i,j} - m_{i+1,j})^2 + (m_{i,j} - m_{i,j-1})^2 + (m_{i,j} - m_{i,j+1})^2$$

The CMI measures how the co-mutation strength of one residue pair differs from its neighboring pairs. By using CMI instead of MI, we can remove the background bias of MI in a local region, as shown in Figure 3-2-2. In the case that there are only a small number of sequence homologs available, some conserved positions, which usually have entropy50.3, may have very low MI, which may result in artificially high CMI. To avoid this, we directly set the CMI of these positions to 0.

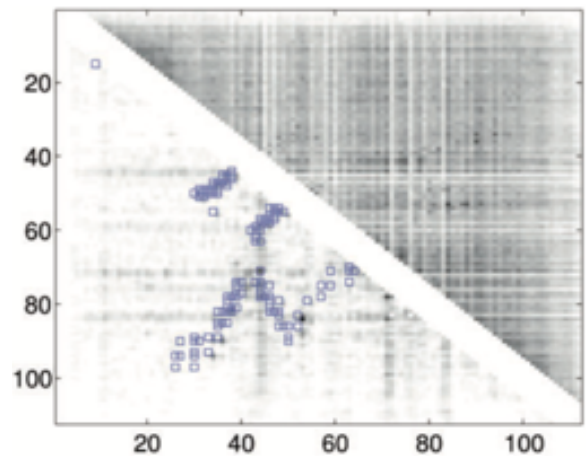

**Figure 3-2-2. The CMI (lower triangle) and MI (upper triangle) of protein 1j8bA. The native contact pairs are marked by boxes.**

## The integer linear programming method

We use binary variables to define the contact map. Let $I$ and $j$ denote residue positions and $L$ denote the protein length. Let $u$ and $v$ index secondary structure segments of a protein. Let $Begin(u)$ and $End(u)$ denote the first and last residues of the segment u and $Sseg(u)$ the set $\{i|Begin(u) \leq i \leq End(u)\}$. Let $Sstype(u)$ denote the secondary structure type of one residue or one segment $u$. Let $Len(u)$ denote the length of the segment $u$. We use the binary variables in Table 3-2-1.

**Table 3-2-1. The binary variables used in the ILP formulation.**

| Variables | Explanations |
|---|---|
| $X_{i,j}$ | Equal to 1 if there is a contact between two residues i and j. |
| $AP_{u,v}$ | Equal to 1 if two -strands u and v form an anti-parallel -sheet. |
| $P_{u,v}$ | Equal to 1 if two -strands u and v form a parallel -sheet. |
| $S_{u,v}$ | Equal to 1 if two -strands u and v form a -sheet. |
| $T_{u,v}$ | Equal to 1 if there is an -bridge between two helices u and v. |
| $R_r$ | A non-negative integral relaxation variable of the $r$th soft constraint. |

**Table 3-2-2. The empirical values of $b_{s1,s2}$ calculated from the training data. The first column indicates the combination of two secondary structure types: H ( -helix), E ( -strand) or C (coil). Each row contains two statistical values for a pair of secondary structure types. Column '95%': 95% of the secondary structure pairs have the number of contacts smaller than the value in this column; column 'Max': the largest number of contacts**

| s1,s2 | 95% | Max |
|---|---|---|
| H,H | 5 | 12 |
| H,E | 3 | 10 |
| H,C | 4 | 11 |
| E,H | 4 | 12 |
| E,E | 9 | 13 |
| E,C | 6 | 15 |
| C,H | 3 | 12 |
| C,E | 5 | 12 |
| C,C | 6 | 20 |

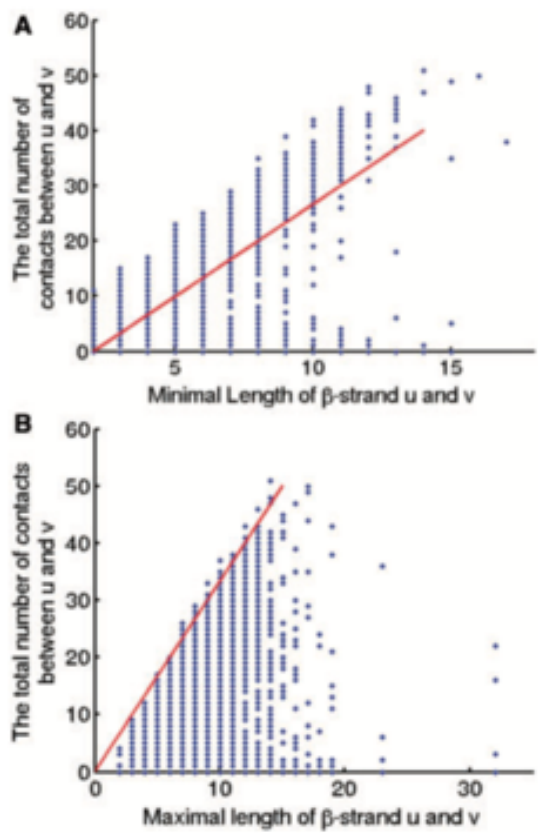

**Figure 3-2-3. The skew lines indicate the bounds for the total number of contacts between two beta-strands. (A) Lower bound; (B) upper bound.**

The objective function is defined on the contacts. Intuitively, we shall choose those contacts with the highest probability predicted by our RF model, i.e. the objective function shall be the sum of predicted probabilities of the selected contacts. However, the selected contacts shall also minimize the violation of the physical constraints. To enforce this, we use a vector of relaxation variables R to measure the degree to which all the soft constraints are violated. Thus, our objective function is as follows.

$$\max_{X,R} \sum_{j-i \geq 6} \left( X_{i,j} \times P_{i,j} \right) - g(R)$$

In the formula above, $P_{i,j}$ is the contact probability predicted by our RF model for two residues and $g(R) = \sum R_r$ is a linear penalty function where *r* runs over all the soft constraints. The relaxation variables will be further explained in the following section.

The constraints of the linear integer programming contain both soft and hard constraints. There is a single relaxation variable for each group of soft constraint, but the hard constraints are strictly enforced. We penalize the violation of soft constraints by incorporating the relaxation variables to the objective function. The constraints in Groups 1, 2 and 6 are soft constraints. Those in Groups 3, 4, 5 and 7 are hard constraints, some of which are similar to what are used by Astro-Fold (Klepeis and Floudas, 2003).

Constraint group 1. This group of soft constraints bound from above the total number of contacts that can be formed by a single residue *i* (in a secondary structure type s1) with all the other residues in a secondary structure type s2.

$$\forall i: SStype(i) = s1, \sum_{j:SStype(j)=s2} X_{i,j} \leq R_1 + b_{s1,s2}$$

In the formula above, $b_{s1,s2}$ is a constant empirically determined from our training data (Table 3-2-2), and $R_1$ is the relaxation variable.

Constraint group 2. This group of constraints bound the total number of contacts between two strands sharing at least one contact. Let u and v denote two $\beta$-strands. We have the following equations.

$$\sum_{i \in SSeg(v), j \in SSeg(u)} X_{i,j} + R_2 \geq 3 \times S_{u,v} \times \min(Len(u), Len(v))$$

$$\sum_{i \in SSeg(v), j \in SSeg(u)} X_{i,j} + R_2 \geq 3.3 \times \max(Len(u), Len(v)) + R_3$$

The two constraints are explained in Figure 3-2-3 as follows. Figure 3-2-3A shows that the total number of contacts between two -strands diverges into two groups when *min(Len(u), Len(v))* ≤ 9. One group is due to $\beta$-strand pairs forming a $\beta$-sheet. The other may be due to incorrectly predicted $\beta$-strands or $\beta$-strand pairs not in a $\beta$-sheet. Figure 3-2-3B shows that the total number of contacts between a pair of $\beta$-strands has an upper bound proportional to the length of the longer $\beta$-strand. As there are points below the skew line in Figure 3-2-3A, which indicate that a $\beta$-strand pair may have fewer than 3 times of *min(Len(u), Len(v))* contacts, we add a relaxation variable $R_2$ to the lower bound constraints in Group 2. Similarly, we use a relaxation variable $R_3$ for the upper bound constraints.

Group 3. When two strands form an anti-parallel beta-sheet, the contacts of neighboring residue pairs shall satisfy the following constraints.

$$X_{i,j} \geq X_{i-1,j+1} + X_{i+1,j-1} - 1$$

In the above formula, *i, i+1, i-1* are residues in one strand, and *j, j+1, j-1* are residues in the other strand.

Group 4. When there are parallel contacts between two strands, the contacts of neighboring residue pairs should satisfy the following constraints.

$$X_{i,j} \geq X_{i-1,j-1} + X_{i+1,j+1} - 1$$

In the formula above, *i, i±1* are residues in one strand, and *j, j±1* are residues in the other strand.

Group 5. One $\beta$-strand can form $\beta$-sheets with up to two other $\beta$-strands.

$$\sum_{v:SStype(v)=beta} S_{u,v} \leq 2$$

Group 6. There is no contact between the start and end residues of a loop segment u.

$$X_{i,j} \leq 0 + R_4, i = Begin(u), j = End(u)$$

In our training set, there are totally around 8000 loop segments, and only 3.4% of them have a contact between the start and end residues. To allow the rare cases, we use a relaxation variable $R_4$ in the constraints.

Group 7. One residue, *i*, cannot have contacts with both segments, *j* and *j + 2* when *j* and *j + 2* are in the same helix.

$$X_{i,j} + X_{i,j+2} \leq 1$$

Group 8. This group of constraints models the relationship among different groups of variables.

$$AP_{u,v} + P_{u,v} = S_{u,v}$$

$$X_{i,j} \leq S_{u,v}, \forall i \in SSeg(u), j \in SSeg(v)$$

$$\sum_{i \leq i < j \leq L, j-i \geq 6} X_{i,j} = k$$

In the above formula, *k* is the number of top contacts we want to predict.

Our ILP model is solved by IBM CPLEX academic version V12.1 (CPLEX, 2009).

Training data. It consists of 900 non-redundant protein structures, any two of which share no more than 25% sequence identity. As there are far fewer contacting residue pairs than non-contacting pairs, we use all the contacting pairs and randomly sample only 20% of the non-contacting pairs as the training data. All the training proteins are selected before CASP10 (the 10th Critical Assessment of Structure Prediction) started in May 2012.

Test data I: CASP10. This set contains 123 CASP10 targets with publicly available native structures. Meanwhile, 36 of them are classified as hard targets because the top half of their submitted models have average TM-score less than 0.5. When they were just released, most of the CASP10 targets share low sequence identity (less than 25%) with the proteins in PDB. BLAST indicates that there are only five short CASP10 targets (about 50 residues), which have sequence identity slightly larger than 30% with our training proteins.

Test data II: Set600. This set contains 601 proteins randomly extracted from PDB25 (Brenner et al., 2000) and was constructed before CASP10 started. The test proteins have the following properties: (i) they share less than 25% sequence identity with the training proteins; (ii) all proteins have at least 50 residues and an X-ray structure with resolution better than 1.9 angstrom; and (iii) all the proteins have at least five residues with predicted secondary structure being alpha-helix or beta-strand.

Both the training set and Set600 are sampled from PDB25 (Wang and Dunbrack, 2003), in which any two proteins share ≤25% sequence identity. Sequence identity is calculated using the method in (Brenner et al., 2000).

Programs to be compared. We compare our method, denoted as PhyCMAP, with four state-of-the-art methods: CMAPpro (Di Lena et al., 2012), NNcon (Tegge et al., 2009), PSICOV (Jones et al., 2012) and Evfold (Morcos et al., 2011). We run NNcon, PSICOV and Evfold locally and CMAPpro through its web server. We do not compare our method with Astro-Fold because it is not publicly available. Further, it does not perform well because of lack of evolutionary information.

Evaluation methods. Depending on the chain distance of the two residues, there are three kinds of contacts: short-range, medium-range and long-range. Short-range contacts are closely related to local conformation and are relatively easy to predict.

Medium-range and long-range contacts determine the overall shape of a protein and are more challenging to predict. We evaluate prediction accuracy using the top 5, L/10, L/5 predicted contacts, where L is the protein length.

$M_{eff}$: the number of non-redundant sequence homologs. Given a target and the multiple sequence alignment of all of its homologs, we measure the number of non-redundant sequence homologs by $M_{eff}$ as follows.

$$M_{eff} = \sum_i \frac{1}{\sum_j S_{i,j}}$$

In the above formula, both i and j go over all the sequence homologs, and $S_{i,j}$ is a binary similarity value between two proteins. Following Evfold (Morcos et al., 2011), we compute the similarity of two sequence homologs using their hamming distance. That is, $S_{i,j}$ is 1 if the normalized hamming distance is less than 0.3, 0 otherwise.

### *Results*

**Performance on the CASP10 set.** As shown in Table 3-2-3, on the whole-CASP10 set, our PhyCMAP significantly exceeds the second best method CMAPpro in terms of the accuracy of the top five, L/10 and L/5 predicted contacts. The advantage of PhyCMAP over CMAPpro becomes smaller but still substantial when short-range contacts are excluded from consideration. PhyCMAP significantly outperforms NNcon, PSICOV and Evfold no matter how the performance is evaluated.

**Table 3-2-3. This table lists the prediction accuracy of PhyCMAP, PSICOV, NNcon, CMAPpro and Evfold on short-, medium- and long-range contacts, tested on CASP10 (123 targets)**

| Method | Short-range, sequence distance from 6 to 12 | Medium- and long-range, sequence distance >12 | Medium-range, sequence distance >12 and ≤24 | Long-range, sequence distance >24 |
|---|---|---|---|---|

|  | Top 5 | L/10 | L/5 | Top 5 | L/10 | L/5 | Top 5 | L/10 | L/5 | Top 5 | L/10 | L/5 |
| --- | --- | --- | --- | --- | --- | --- | --- | --- | --- | --- | --- | --- |
| PhyCMAP ($C_\alpha$) | 0.623 | 0.555 | 0.459 | 0.650 | 0.584 | 0.523 | 0.585 | 0.518 | 0.448 | 0.421 | 0.363 | 0.320 |
| PhyCMAP ($C_\beta$) | 0.667 | 0.580 | 0.482 | 0.655 | 0.604 | 0.539 | 0.621 | 0.550 | 0.465 | 0.514 | 0.425 | 0.373 |
| PSICOV ($C_\alpha$) | 0.294 | 0.225 | 0.179 | 0.397 | 0.345 | 0.306 | 0.384 | 0.303 | 0.255 | 0.350 | 0.277 | 0.226 |
| PSICOV ($C_\beta$) | 0.379 | 0.281 | 0.223 | 0.522 | 0.458 | 0.405 | 0.515 | 0.387 | 0.316 | 0.457 | 0.372 | 0.315 |
| NNcon ($C_\alpha$) | 0.595 | 0.499 | 0.399 | 0.472 | 0.409 | 0.358 | 0.463 | 0.393 | 0.334 | 0.286 | 0.239 | 0.188 |
| CMAPpro ($C_\alpha$) | 0.506 | 0.437 | 0.368 | 0.517 | 0.466 | 0.424 | 0.485 | 0.414 | 0.363 | 0.380 | 0.336 | 0.297 |
| CMAPpro ($C_\beta$) | 0.543 | 0.477 | 0.395 | 0.519 | 0.466 | 0.415 | 0.491 | 0.419 | 0.370 | 0.395 | 0.347 | 0.313 |
| Evfold ($C_\alpha$) | 0.236 | 0.193 | 0.165 | 0.380 | 0.326 | 0.295 | 0.351 | 0.294 | 0.249 | 0.328 | 0.257 | 0.225 |

**Performance on the 36 hard CASP10 targets.** As shown in Table 3-2-4, on the 36 hard CASP10 targets, our PhyCMAP exceeds the second best method CMAPpro by 5–7% in terms of the accuracy of the top five, L/10 and L/5 predicted contacts. The advantage of PhyCMAP over CMAPpro becomes smaller but still substantial when short-range contacts are excluded from consideration. PhyCMAP significantly outperforms NNcon, PSICOV and Evfold no matter how many predicted contacts are evaluated. PSICOV and Evfold almost fail on these hard CASP10 targets. By contrast, CMAPpro, NNcon and PhyCMAP still work, although they do not perform as well as on the whole CASP10 set.

**Table 3-2-4. Prediction accuracy on the 36 hard CASP10 targets.**

| Method | Short-range, sequence distance from 6 to 12 | | | Medium and long-range, sequence distance >12 | | |
| --- | --- | --- | --- | --- | --- | --- |
|  | Top 5 | L/10 | L/5 | Top 5 | L/10 | L/5 |
| PhyCMAP ($C_\alpha$) | 0.456 | 0.439 | 0.378 | 0.394 | 0.378 | 0.325 |
| PhyCMAP ($C_\beta$) | 0.478 | 0.469 | 0.414 | 0.444 | 0.409 | 0.363 |
| PSICOV ($C_\alpha$) | 0.100 | 0.083 | 0.082 | 0.183 | 0.156 | 0.150 |
| PSICOV ($C_\beta$) | 0.144 | 0.113 | 0.103 | 0.239 | 0.196 | 0.180 |
| NNcon ($C_\alpha$) | 0.400 | 0.372 | 0.320 | 0.367 | 0.317 | 0.289 |
| CMAPpro ($C_\alpha$) | 0.383 | 0.347 | 0.314 | 0.328 | 0.322 | 0.292 |
| CMAPpro ($C_\beta$) | 0.433 | 0.398 | 0.344 | 0.394 | 0.362 | 0.308 |
| Evfold ($C_\alpha$) | 0.100 | 0.095 | 0.094 | 0.194 | 0.179 | 0.156 |

Note that both PSICOV and Evfold, two recent methods receiving a lot of attentions from the community, do not perform well on the CASP10 set. This is partially because they require a large number of sequence homologs for the protein under prediction. Nevertheless, most of the CASP targets, especially the hard ones, do not have so many sequence homologs because a protein with so many homologs likely has similar templates in PDB and thus, were not used by CASP.

We also studies the relationship between prediction accuracy and the number of sequence homologs. We divide the 123 CASP10 targets into five groups according to their log$M_{eff}$ values: (0,2), (2,4), (4,6), (6,8), (8,10), which contain 19, 17, 25, 36 and 26 targets, respectively. Meanwhile, $M_{eff}$ is the number of non-redundant sequence homologs for the protein under consideration. Only medium- and long-range contacts are considered here. Figure 3-2-4 clearly shows that the prediction accuracy increases with respect to $M_{eff}$. The more non-redundant not substantial for C contacts prediction when short-range contacts are excluded from consideration. PhyCMAP also outperforms NNcon and CMAPpro on this set. As shown in Table 3-2-5, on the second subset, PhyCMAP significantly outperforms PSICOV and is slightly better than CMAPpro and NNcon. These results again confirm that our method applies to a protein without many sequence homologs, on which PSICOV and Evfold usually fail.

**Table 3-2-5. Benchmark on the 471 proteins with $M_{eff}$ larger than 100.**

| Method | Short-range, sequence distance from 6 to 12 | | | Medium- and long-range, sequence distance >12 | | |
|---|---|---|---|---|---|---|
| | Top 5 | L/10 | L/5 | Top 5 | L/10 | L/5 |
| PhyCMAP ($C_α$) | 0.761 | 0.653 | 0.539 | 0.716 | 0.675 | 0.611 |
| PhyCMAP ($C_β$) | 0.746 | 0.637 | 0.531 | 0.731 | 0.656 | 0.608 |
| PSICOV ($C_α$) | 0.457 | 0.341 | 0.257 | 0.528 | 0.465 | 0.411 |
| PSICOV ($C_β$) | 0.584 | 0.425 | 0.316 | 0.732 | 0.646 | 0.565 |
| NNcon ($C_α$) | 0.512 | 0.432 | 0.355 | 0.432 | 0.361 | 0.308 |
| CMAPpro ($C_α$) | 0.682 | 0.551 | 0.431 | 0.710 | 0.642 | 0.574 |

| Method | Short-range, sequence distance from 6 to 12 | | | Medium- and long-range, sequence distance >12 | | |
|---|---|---|---|---|---|---|
| | Top 5 | L/10 | L/5 | Top 5 | L/10 | L/5 |
| CMAPpro ($C_\beta$) | 0.671 | 0.542 | 0.436 | 0.674 | 0.601 | 0.532 |
| Evfold ($C_\alpha$) | 0.379 | 0.297 | 0.234 | 0.473 | 0.438 | 0.400 |

PhyCMAP also outperforms homologs are available, the better prediction accuracy can be achieved by PhyCMAP, Evfold and PSICOV. However, CMAPpro and NNcon have decreased accuracy when log$M_{eff}$ less than 8.

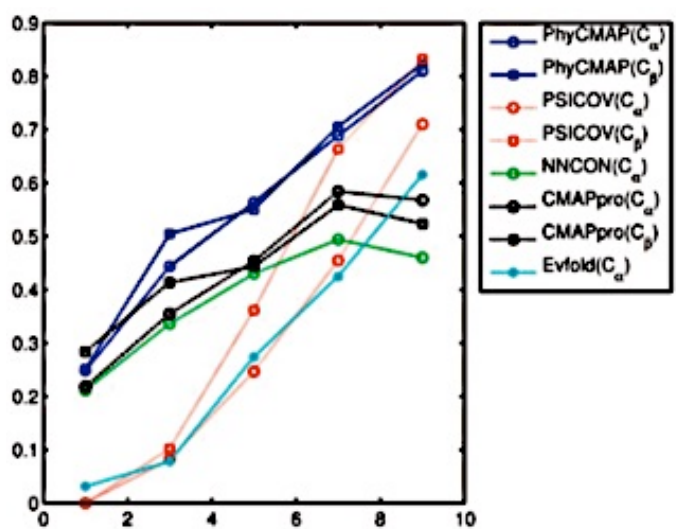

**Figure 3-2-4. The relationship between prediction accuracy and the number of non-redundant sequence homologs (Meff). x-axis is logMeff and y-axis is the mean accuracy of top L/10 predicted contacts on the corresponding CASP10 target group. Only medium- and long-range contacts are considered**

Figure 3-2-4 also shows that PhyCMAP outperforms Evfold, CMAPpro and NNcon across all $M_{eff}$. PhyCMAP greatly outperforms PSICOV in predicting $C_\alpha$ contacts regardless of $M_{eff}$ and also in predicting $C_\beta$ contacts when log$M_{eff}$ less than 6. PhyCMAP has comparable performance as PSICOV in predicting $C_\beta$ contacts when log$M_{eff}$ less than 6.

**Performance on Set600.** To fairly compare our method with Evfold (Morcos et al., 2011) and PSICOV (Jones et al., 2012), both of which require a large number of sequence homologs, we divide Set600 into two subsets based on the amount of homologous information available for the protein under prediction. The first subset is relatively easier, containing 471 proteins with $M_{eff}$ less than 100 (see Section 2 for definition). All the proteins in this subset have more than 500 sequence homologs, which satisfies the requirement of PSICOV. The second subset is more challenging to predict, containing 130 proteins with $M_{eff}$ less than 100. As shown in Table 3-2-6, even on the first subset, PhyCMAP still exceeds PSICOV and Evfold, although the advantage over PSICOV is not substantial for $C_\beta$ contacts prediction when short-range contacts are excluded from consideration. PhyCMAP also outperforms NNcon and CMAPpro on this set. As shown in Table 3-2-6, on the second subset, PhyCMAP significantly outperforms PSICOV and is slightly better than CMAPpro and NNcon. These results again confirm that our method applies to a protein without many sequence homologs, on which PSICOV and Evfold usually fail.

**Table 3-2-6. Benchmark on the 130 proteins with $M_{eff} \leq 100$.**

| Method | Short-range, sequence distance from 6 to 12 | | | Medium- and long-range, sequence distance >12 | | |
|---|---|---|---|---|---|---|
| | Top 5 | L/10 | L/5 | Top 5 | L/10 | L/5 |
| PhyCMAP ($C_\alpha$) | 0.534 | 0.451 | 0.377 | 0.432 | 0.372 | 0.319 |
| PhyCMAP ($C_\beta$) | 0.505 | 0.435 | 0.365 | 0.418 | 0.364 | 0.314 |
| PSICOV ($C_\alpha$) | 0.060 | 0.061 | 0.064 | 0.049 | 0.039 | 0.035 |
| PSICOV ($C_\beta$) | 0.077 | 0.070 | 0.073 | 0.069 | 0.050 | 0.045 |
| NNcon ($C_\alpha$) | 0.442 | 0.363 | 0.309 | 0.368 | 0.339 | 0.301 |
| CMAPpro ($C_\alpha$) | 0.435 | 0.365 | 0.314 | 0.368 | 0.331 | 0.300 |
| CMAPpro ($C_\beta$) | 0.532 | 0.434 | 0.353 | 0.358 | 0.331 | 0.280 |

It should be noticed that CMAPpro used Astral 1.73 (Brenner et al., 2000; Di Lena et al., 2012) as its training set, which shares more than 90% sequence identity with 226 proteins in Set600 (180 with $M_{eff}$ larger than 100 and 46 with $M_{eff}$ less than 100). To more fairly compare the prediction methods, we exclude the 226 proteins from Set600 that share more than 90% sequence identity with the CMAPpro training set. Here, the sequence identity is calculated using CD-HIT (Li and Godzik, 2006; Li et al., 2001). This results in a set of 291 proteins with $M_{eff}$ larger than 100 and 84 proteins $M_{eff}$ less than 100. Table 3-2-7 shows that PhyCMAP greatly outperforms CMAPpro and Evfold on the reduced dataset. PhyCMAP also outperforms PSICOV in predicting C-alpha contacts, but it is slightly worse in predicting long-range C-beta contacts.

Table 3-2-7. This table lists the prediction accuracy of PhyCMAP, PSICOV, NNcon, CMAPpro and Evfold on short-, medium- and long-range contacts, tested on Set600.

| Method | Short-range, sequence distance from 6 to 12 | | | Medium- and long-range, sequence distance >12 | | | Medium-range, sequence distance >12 and ≤24 | | | Long-range, sequence distance >24 | | |
|---|---|---|---|---|---|---|---|---|---|---|---|---|
| | Top 5 | L/10 | L/5 | Top 5 | L/10 | L/5 | Top 5 | L/10 | L/5 | Top 5 | L/10 | L/5 |
| a) The 291 proteins in Set600 with Meff >100 and ≤90% sequence identify with Astral 1.73 | | | | | | | | | | | | |
| PhyCMAP($C_\alpha$) | 0.759 | 0.653 | 0.536 | 0.713 | 0.680 | 0.622 | 0.639 | 0.570 | 0.496 | 0.582 | 0.528 | 0.461 |
| PhyCMAP($C_\beta$) | 0.741 | 0.641 | 0.534 | 0.746 | 0.653 | 0.611 | 0.655 | 0.571 | 0.500 | 0.636 | 0.550 | 0.477 |
| PSICOV($C_\alpha$) | 0.459 | 0.343 | 0.258 | 0.528 | 0.469 | 0.417 | 0.462 | 0.363 | 0.282 | 0.483 | 0.418 | 0.358 |
| PSICOV($C_\beta$) | 0.582 | 0.422 | 0.314 | 0.733 | 0.650 | 0.569 | 0.647 | 0.496 | 0.371 | 0.674 | 0.584 | 0.495 |
| NNcon($C_\alpha$) | 0.475 | 0.390 | 0.318 | 0.377 | 0.313 | 0.267 | 0.342 | 0.284 | 0.236 | 0.224 | 0.182 | 0.152 |
| CMAPpro($C_\alpha$) | 0.643 | 0.519 | 0.412 | 0.689 | 0.618 | 0.554 | 0.580 | 0.511 | 0.439 | 0.527 | 0.469 | 0.416 |
| CMAPpro($C_\beta$) | 0.642 | 0.520 | 0.422 | 0.653 | 0.580 | 0.515 | 0.573 | 0.494 | 0.421 | 0.504 | 0.444 | 0.396 |
| Evfold($C_\alpha$) | 0.382 | 0.297 | 0.235 | 0.488 | 0.442 | 0.398 | 0.451 | 0.366 | 0.289 | 0.442 | 0.389 | 0.342 |
| b) The 84 proteins in Set600 with Meff ≤100 and ≤90% sequence identity with Astral 1.73 | | | | | | | | | | | | |
| PhyCMAP($C_\alpha$) | 0.580 | 0.488 | 0.404 | 0.481 | 0.430 | 0.357 | 0.476 | 0.417 | 0.335 | 0.204 | 0.179 | 0.159 |
| PhyCMAP($C_\beta$) | 0.548 | 0.468 | 0.392 | 0.454 | 0.408 | 0.345 | 0.452 | 0.399 | 0.331 | 0.220 | 0.214 | 0.187 |
| PSICOV($C_\alpha$) | 0.070 | 0.071 | 0.072 | 0.065 | 0.050 | 0.044 | 0.074 | 0.055 | 0.049 | 0.063 | 0.043 | 0.035 |
| PSICOV($C_\beta$) | 0.081 | 0.078 | 0.083 | 0.088 | 0.068 | 0.059 | 0.092 | 0.066 | 0.059 | 0.076 | 0.058 | 0.046 |
| | 0.535 | 0.421 | 0.342 | 0.324 | 0.298 | 0.248 | 0.348 | 0.321 | 0.271 | 0.162 | 0.132 | 0.114 |

| Method | Short-range, sequence distance from 6 to 12 | | | Medium- and long-range, sequence distance >12 | | | Medium-range, sequence distance >12 and ≤24 | | | Long-range, sequence distance >24 | | |
|---|---|---|---|---|---|---|---|---|---|---|---|---|
| | Top 5 | L/10 | L/5 | Top 5 | L/10 | L/5 | Top 5 | L/10 | L/5 | Top 5 | L/10 | L/5 |
| NNcon($C_\alpha$) | | | | | | | | | | | | |
| CMAPpro($C_\alpha$) | 0.465 | 0.370 | 0.316 | 0.346 | 0.328 | 0.285 | 0.360 | 0.332 | 0.286 | 0.173 | 0.169 | 0.158 |
| CMAPpro($C_\beta$) | 0.447 | 0.367 | 0.321 | 0.346 | 0.320 | 0.287 | 0.366 | 0.331 | 0.290 | 0.191 | 0.189 | 0.176 |
| Evfold($C_\alpha$) | 0.074 | 0.068 | 0.066 | 0.079 | 0.058 | 0.039 | 0.074 | 0.053 | 0.045 | 0.063 | 0.042 | 0.032 |

*Contribution of contrastive mutual information and pairwise contact scores*

The CMI and HPS help improve the performance of our RF model. Table 3-2-8 shows their contribution to the prediction accuracy on the 471 proteins (with $M_{eff}$ larger than 100) in Set600.

**Table 3-2-8. The contribution of CMI and homologous pair contact scores to $C_\beta$ contact prediction.**

| Method | Short-range contacts | | | Medium- and long-range | | |
|---|---|---|---|---|---|---|
| | Top 5 | L/10 | L/5 | Top 5 | L/10 | L/5 |
| with CMI and HPS | 0.754 | 0.632 | 0.521 | 0.720 | 0.649 | 0.589 |
| no CMI and HPS | 0.600 | 0.570 | 0.487 | 0.538 | 0.560 | 0.506 |

*Contribution of physical constraints*

Table 3-2-9 shows the improvement resulting from the physical constraints (i.e. the ILP method) over the RF method on Set600. On the 471 proteins with $M_{eff}$ larger than 100, ILP improves medium and long-range contact prediction, but not short-range contact prediction. This result confirms that the physical constraints used by our ILP method indeed capture some global properties of a protein contact map. The improvement resulting from the physical constraints is larger on the 130 proteins with $M_{eff}$ less than 100. In particular, the improvement on short-range contacts is substantial. These results may imply that when homologous information is sufficient, we can predict short-range contacts accurately and thus, cannot further

improve the prediction by using the physical constraints. When homologous information is insufficient for accurate contact prediction, we can improve the prediction using the physical constraints, which are complementary to evolutionary information.

Table 3-2-9. The contribution of physical constraints.

| Method | Short-range contacts | | | Medium- and long-range | | |
|---|---|---|---|---|---|---|
| | Top 5 | L/10 | L/5 | Top 5 | L/10 | L/5 |
| 471 proteins in Set600 with $M_{eff}$ >100 | | | | | | |
| RF + ILP | 0.746 | 0.637 | 0.531 | 0.731 | 0.656 | 0.608 |
| RF | 0.754 | 0.632 | 0.521 | 0.720 | 0.649 | 0.589 |
| 130 proteins in Set600 with $M_{eff}$ ≤ 100 | | | | | | |
| RF + ILP | 0.505 | 0.435 | 0.365 | 0.418 | 0.364 | 0.314 |
| RF | 0.445 | 0.368 | 0.299 | 0.414 | 0.342 | 0.281 |

***Specific examples***

We show the contact prediction results of two CASP10 targets T0693D2 and T0677D2 in Figures 3-2-5 and Figure 3-2-6, respectively. T0693D2 has many sequence homologs with $M_{eff}$=2208.39. As shown in Figure 3-2-5, PhyCMAP does well in predicting the long-distance contacts around the residue pair (20,100). For this target, PhyCMAP and Evfold obtain top L/10 prediction accuracy of 0.810 and 0.619, respectively, on medium- and long-range contacts. T0677D2 does not have many sequence homologs with $M_{eff}$=31.53. As shown in Figure 3-2-6, our prediction matches well with the native contacts. PhyCMAP has top L/10 prediction accuracy 0.429 on medium- and long-range contacts, whereas Evfold cannot correctly predict any contacts.

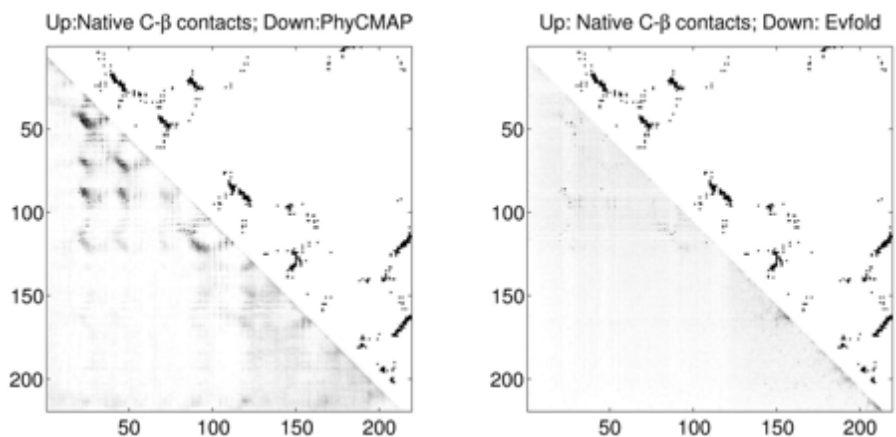

**Figure 3-2-5. The predicted medium- and long-range contacts for T0693D2. The upper triangles display the native $C_β$ contacts. The lower triangles of the left and right plots display the contact probabilities predicted by PhyCMAP and Evfold, respectively.**

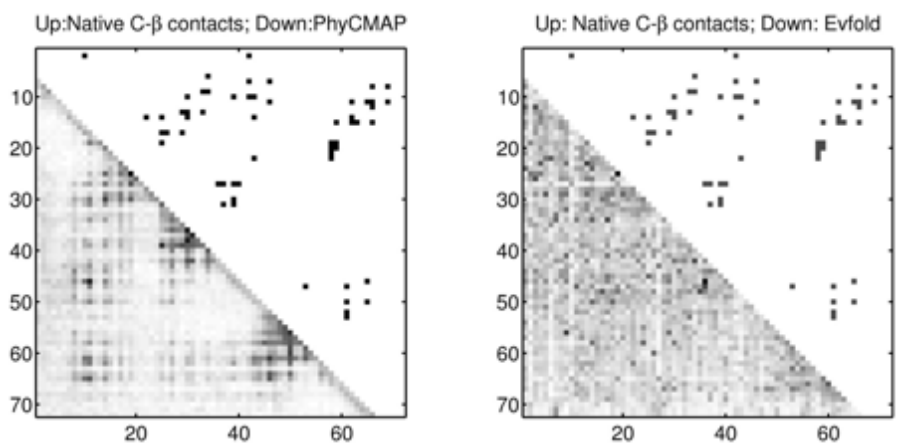

**Figure. 3-2-6. The predicted medium- and long-range contacts for T0677D2. The upper triangles display the native $C_β$ contacts. The lower triangles of the left and right plots display the contact probabilities predicted by PhyCMAP and Evfold, respectively.**

## *Conclusions*

This article has presented a novel method for protein contact map prediction by integrating both evolutionary and physical constraints using machine learning and ILP. Our method differs from currently popular contact prediction methods in that we enforce a few physical constraints, which imply the sparsity constraint (used by PSICOV and Evfold), to the whole-contact map and take into consideration contact correlation. Our method also differs from the first-principle method (e.g. Astro-Fold) in that we exploit evolutionary information from several aspects (e.g. MI, context-specific distance potential and sequence profile) to significantly improve prediction accuracy. Experimental results confirm that our method outperforms existing popular machine- learning methods (e.g. CMAPpro and NNcon) and two recent co-mutation–based methods PSICOV and Evfold regardless of the number of sequence homologs available for the protein under consideration.

The study of our method indicates that the physical constraints are helpful for contact prediction, especially when the protein under consideration does not have many sequence homologs. Nevertheless, the physical constraints exploited by our current method do not cover all the properties of a protein contact map. To further improve prediction accuracy on medium- and long-range contact prediction, we may take into consideration global properties of a protein distance matrix. For example, the pairwise distances of any three residues shall satisfy the triangle inequality. Some residues also have correlated pairwise distances. To enforce this kind of distance constraints, we shall introduce distance variables and also define their relationship with contact variables. By introducing distance variables, we may also optimize the distance probability, as opposed to the contact probability used by our current ILP method. Further, we can also introduce variables of beta-sheet (group of beta-strands) to capture more global properties of a contact map.

One may ask how our approach compares with a model-based filtering strategy in which 3D models are built based on initial predicted contacts and then used to filter incorrect predictions. Our method differs from this general 'model-based filtering' strategy in a couple of aspects. First, it is time-consuming to build thousands or at least hundreds of 3D models with initial predicted contacts. In contrast, our method can do contact prediction (using physical constraints) within minutes. Second, the quality of the 3D models is also determined by other factors, such as energy function and energy optimization (or conformation sampling) methods, whereas our method is independent of these factors. Even if the energy function is accurate, the energy optimization algorithm often is trapped to local minima because the energy function is not rugged. That is, the 3D models resulting from energy minimization are biased toward a specific region of the conformation space, unless an extremely large-scale of conformation sampling is conducted. Therefore, the predicted contacts derived from these models may also suffer from this 'local minima' issue. By contrast, our integer programming method can search through the whole conformation space and find the global optimal solution; thus, it is not biased to any local minima region. By using our predicted contacts as constraints, we may pinpoint to the good region of a conformation space (without being biased by local minima), reduce the search space and significantly speed-up conformation search.

# Chapter 4. 3D structure prediction

## 4.1 RNA 3D structure

In the world of RNA, the secondary structure has a different meaning than in the protein world. RNA secondary structure describes both the short and long range of a RNA 3D structure. It is similar to a protein contact map, but not similar to a protein secondary structure. Compared with 20 amino acids in proteins, there are only 4 nucleotides in RNA sequences, and each nucleotide has its own strong preference for having a contact with the other nucleotide type. The RNA secondary structure prediction problem has been studied for a long time (Akutsu, 2000; Bindewald and Shapiro, 2006; Chen, et al., 2008; Do, et al., 2006; Eddy and Durbin, 1994; Hofacker, 2003; Knudsen and Hein, 2003; Zuker, 2003; Zuker and Sankoff, 1984).

There are not so many RNA 3D prediction methods have been published, including molecular dynamics and knowledge-based methods. Among these are FARNA (Das and Baker, 2007), MC-Sym (Parisien and Major, 2008), and BARNACLE (Frellsen, et al., 2009). In our study, the results showed that a predicted secondary structure can significantly help to improve the sampling efficiency of a RNA 3D structure prediction method (Wang and Xu, 2011). With predicted secondary structures, our method produces more decoys of a high quality compared with other state-of-the-arts methods.

Different from other methods, our method takes input of the predicted secondary structure to guide the 3D structure sampling process. The RNA secondary structure defines the order of sub-sequence sampling. For each sub-sequence, our algorithm samples its 3D structure according to an energy function modelled by a condition random field model.

Predicting RNA tertiary structures is extremely challenging, because of a large conformation space to be explored and lack of an accurate scoring function differentiating the native structure from decoys. The fragment-based conformation sampling method (e.g. FARNA) bears shortcomings that the limited size of a fragment library makes it infeasible to represent all possible conformations well. A recent dynamic Bayesian network method, BARNACLE, overcomes the issue of fragment assembly. In addition, neither of these methods makes use of sequence information in sampling conformations. Here, we present a new probabilistic graphical model, conditional random fields (CRFs), to model RNA sequence–structure relationship, which enables us to accurately estimate the probability of an RNA conformation from sequence. Coupled with a novel tree-guided sampling scheme, our CRF model is then applied to RNA conformation sampling. Experimental results show that our CRF method can model RNA sequence–structure relationship well and sequence information is important for conformation sampling. Our method, named as TreeFolder, generates a much higher percentage of native-like decoys than FARNA and BARNACLE, although we use the same simple energy function as BARNACLE.

## *Method*

### *Representation of an RNA structure and conformation state*

We can represent an RNA 3D structure using a sequence of torsion angles, as shown in Figure 4-1-1. Every nucleotide has in total seven bonds that rotate freely. Six of them lie on the backbone: P–O5 , O5 –C5 , C5 –C4 , C4 –C3 , C3 –O3 and O3 –P. The seventh bond connects a base to atom C1 . As shown in Figure 4-1-2 torsion χ around the seventh bond has a small variance, so we assume that it is independent of the other angles and has a normal distribution. The planar angles between two adjacent bonds on the backbone are almost constants, so are the lengths of the bonds. We use a simplified representation so that we can reduce the number of torsion angles needed for the local conformation of a nucleotide (Cao and Chen, 2005; Duarte and Pyle, 1998; Hershkovitz et al., 2006; Zhang et al., 2008). In particular, we use the torsions τ1 and τ2 on pseudo-bonds P–C4 and C4 –P (see pink lines in Figure 4-1-1). However, to determine coordinates of the six backbone atoms of a nucleotide, we also need two planar angles θ, ψ and another torsion α on bond P–O5 . Overall, we use a five tuple *(τ1, τ2, θ,ψ,α)* to represent the local conformation of a nucleotide. The torsion angles are separated in several groups in the whole angle space Figure 4-1-3. Although there are many different methods to represent an RNA conformation, this simplified representation enables us to rapidly rebuild backbone atoms from angles. Similar representations have also been extensively adopted by previous works (Cao and Chen, 2005; Duarte and Pyle, 1998; Hershkovitz *et al.*, 2006; Zhang *et al.*, 2008).

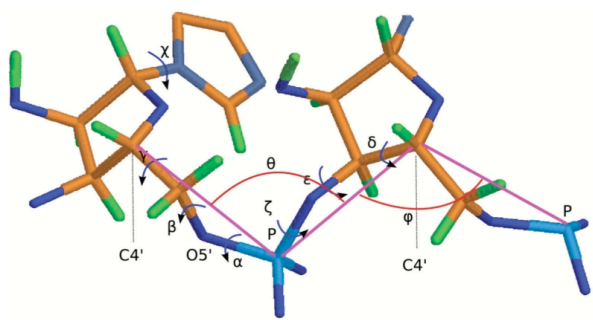

**Figure 4-1-1. Conformation of a nucleotide is represented by angles.**

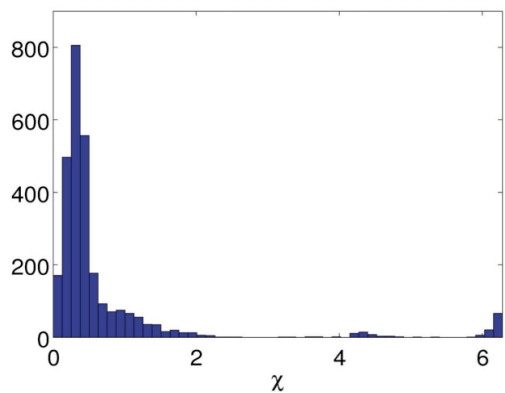

**Figure 4-1-2. Empirical distribution of the torsion angle χ collected from the all representative RNA structures.**

*Our simplified representation does not lose much accuracy*: given the torsion angles, we can rebuild the atom coordinates of an RNA molecule with very good accuracy. As shown in Table 4-1-1, the structures rebuilt from the native angle values (assuming the bond lengt hs are constants) have RMSD <1 Å from their natives.

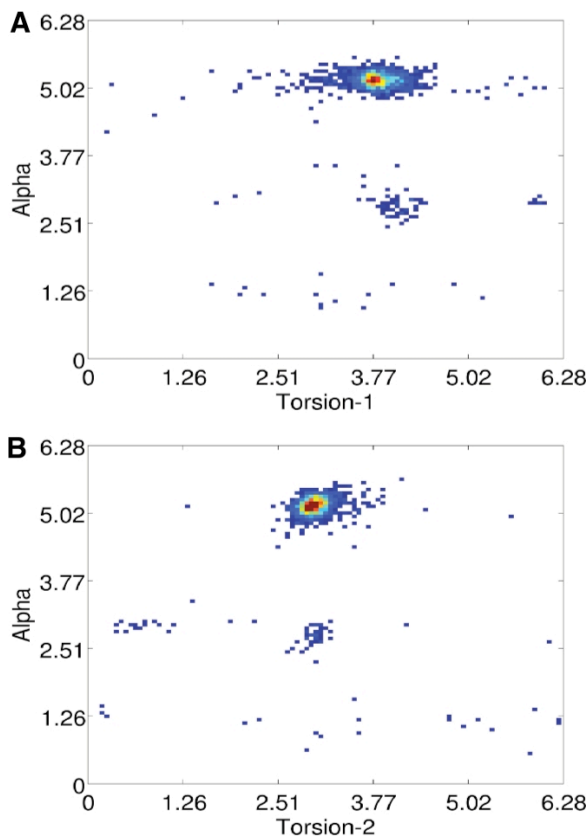

**Figure 4-1-3. (A) Empirical distribution density of the torsion ($\tau_1$) on the pseudo–bond C4′--P and α. (B) Distribution density of the torsion ($\tau_2$) on the pseudo–bond P-*C4*′ and α. The empirical distributions are built from all representative RNA structures.**

*Conformation state*: we use a Gaussian distribution to describe the local conformation preference of one nucleotide. First, we cluster all the angles collected from the experimental structures into dozens of groups (20~100). Then, we calculate the mean and variance in each group and model the angle distribution, using Gaussian distribution. Each group (or cluster) and its Gaussian distribution are identified by an index, which is also denoted as a conformation state. Given the conformation state of a nucleotide, we can sample its real-valued angles from the corresponding distribution. Note that to make angle sampling easy and fast, we assume the torsion angles are independent of one another in Gaussian distribution. Later we will show how to empirically determine the best number of conformation states to achieve the best sampling performance.

### *CRF model for RNA sequence–structure relationship*

Our CRF method can estimate the probability of an RNA conformation from the primary sequence and secondary structure. A CRF model consists of two major components: input features and output labels. The input features at each nucleotide include its nucleotide types, base pairing states and its neighbor nucleotide types. The input features are encoded as a vector of binary variables. The base pairing states can be predicted using some secondary structure prediction programs (Akutsu, 2000; Do *et al.*, 2006; Eddy and Durbin, 1994; Gardner and Giegerich, 2004; Knudsen and Hein, 2003; Mathews and Turner, 2006; Poolsap *et al.*, 2009; Zuker, 2003) with reasonable accuracy. The base pairing information can also be obtained using some experimental methods (Gewirth *et al.*, 1987; Wohnert *et al.*, 1999; Zwahlen *et al.*, 1997), which are much less expensive than those methods determining RNA tertiary structures. The output label at each nucleotide is a conformation state (also called label in CRF). It is also the index of the cluster, which the angles at this nucleotide belongs to.

In contrast to BARNACLE (Frellsen *et al.*, 2009) estimating the generative probability of an RNA structure, our CRF model estimates the conditional probability of an RNA structure, represented as a conformation state vector *y*, from the input feature vector *x* as follows.

$$P(Y = y | X = x) = \frac{1}{Z(x)} \exp[\sum_{i=1}^{L} \psi(y_i, x) + \sum_{i=1}^{L-1} \phi(y_i, y_{i+1})]$$

$$Z(x) = \sum_y \exp\left[\sum_{i=1}^{L} \psi(y_i, x) + \sum_{i=1}^{L-1} \phi(y_i, y_{i+1})\right] \quad (4\text{-}1\text{-}1)$$

$$y = (y_1, \ldots, y_L), \quad \psi(y_i, x) = V_{y_i}^T x, \quad \phi(y_i, y_{i+1}) = W_{y_i, y_{i+1}}$$

In the formula above, $Z(x)$ is the partition function; $x_i$ is the feature vector at position $i$; $y_i$ is the label at position $i$; $W_{i,j}$ is the weight for transition from state $i$–$j$; $V_i$ is the weight factor for predicting state $i$ from an input feature $x$; $L$ is the length of RNA, i.e. the number of nucleotides. The function $\psi$ describes dependency between a conformation state and the input features and thus, called a label feature function. The function describes dependency between two adjacent states and thus called an edge feature function.

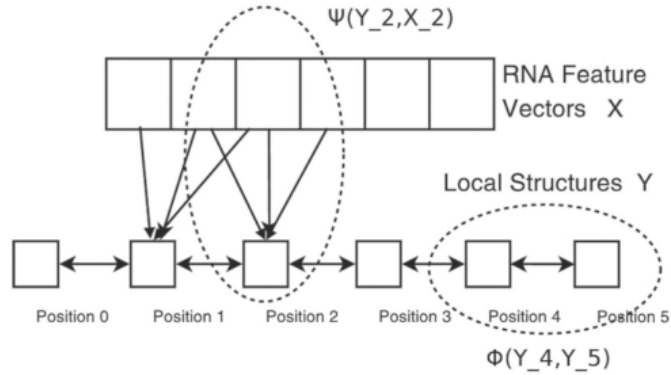

**Figure 4-1-4 A linear-chain CRF model describes the RNA sequence–structure relationship. The input feature vector $X$ contains sequence information and the label (state) vector $Y$ contains local conformation states.**

Figure 4-1-4 shows a linear-chain CRF model for the sequence–structure relationship of an artificial RNA with five nucleotides. We also extend $\psi$ to a linear combination of features of the adjacent nucleotides in a sliding window. That is, $\psi$ is a linear function of $\tilde{x}_i = [x_{i-WL/2} \cdots x_{i+WL/2}]$, $WL$ is the window size to be determined later.

Once the CRF model is trained, we can calculate the (marginal) probability of a conformation state at a given position, using the forward–backward algorithm as follows.

$$P(Y_t = y_t | X = x) = \frac{1}{Z(x)} F(t, y_t, x) B(t+1, y_t, x)$$

$$F(t, y, x) = \begin{cases} \sum_{u=0}^{N} F(t-1, u, x) \exp(\Phi(u, y) + \Psi(y, x_t)), & t > 1 \\ \exp(\Psi(y, x_t)), & t = 1 \end{cases}$$

$$B(t, y, x) = \begin{cases} \sum_{u=0}^{N} B(t+1, u, x) \exp(\Phi(y, u) + \Psi(y, x_{t+1})), & t < L \\ \exp(\Phi(y, u)), & t = L \end{cases}$$

$$Z(x) = \sum_{u=0}^{N} F(L, u, x)$$

We train our CRF model by maximizing the occurring probability of a set of training RNAs with solved structures. In order to avoid overfitting, we also enforce regularization on the model parameters. As such, we train the model parameters by maximizing the following regularized log-likelihood.

$$\log \left( \prod_k P(Y = y^k | X = x^k) \right) + \lambda ||w||_2 + \mu ||V||_2$$

Meanwhile, $y^k$ and $x^k$ are the conformation state vector and input feature vector of the $k$-th training RNA, $W$ and $V$ are model parameters defined in Equation (4-1-1) and λ and μ are the regularization factors. This maximization problem can be solved to optimal using the L-BFGS algorithm (Liu and Nocedal, 1989).

We also extend the first-order CRF model to the second-order model so that we can capture dependency among three adjacent nucleotides. As in Figure 4-1-5, two adjacent positions are combined to a single super node. All the algorithms for the first-order CRF model can be easily extended to the second-order model.

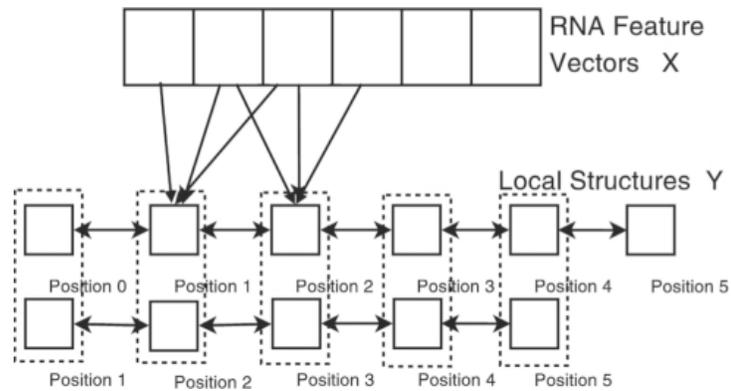

**Figure 4-1-5. The second-order CRF model describes RNA sequence–structure relationship. A super node in this model contains the conformation states in two adjacent positions.**

***A tree-guided conformation sampling algorithm***

Once the CRF model is trained, we can use it to sample conformations for a segment in an RNA molecule. By combining this segment conformation sampling algorithm with a tree representation of the RNA base pairing information, we can have a tree-guided conformation sampling scheme, which enables us to sample conformations for two segments far away from each other along the sequence.

*Building a guide tree for conformation sampling*: the guide tree represents the base pairing information in an RNA, which can be predicted using a secondary structure prediction method or determined by experimental methods. In the case of pseudo-knots, we remove the minimal number of base pairings so that a tree can be built. Since the pseudo-knots do not occur frequently, removal of a small number of base pairings does not impact our method. Note that all the base pairings are taken into consideration in calculating the energy of a sampled conformation. Therefore, removal of some base pairs in tree construction will not impact the formation of pseudo- knots, since we also use energy function to guide the folding simulation.

Given the base pairing information, we build a guide tree as follows. The root node in the tree corresponds to the whole RNA. Given a base pair $(i, j)$, we have one node in the tree corresponding to the segment between $i$ and $j$. One node $A$ is the child of the other node $B$ if and only if the segment corresponding to $B$ is the minimal segment containing the segment corresponding to $A$. In case that one node has more than two child nodes, we can always add some intermediate nodes so that any node has at most two child nodes. For example, supposing node $B$, corresponding to segment $(i, j)$, has three child nodes $A_1$, $A_2$ and $A_3$, where $A_k$ corresponds to segment $(i_k, j_k)$ and $i<i_1<j_1<i_2<j_2<i_3<j_3<j$. We can add an intermediate node $C$ for segment $(i_1, j_2)$ so that $C$ becomes the parent node of $A_1$ and $A_2$ and $B$ has only two child nodes $A_3$ and $C$.

*Segment conformation sampling algorithm*: This sampling algorithm consists of two steps: sampling a label for each nucleotide, in the segment, by the probability calculated from the CRF model and sampling real-valued angles from Gaussian distribution corresponding to a label. We use a forward–backward algorithm to sample the label sequence of a segment from position $i$ to $j$. The algorithm iteratively draws a conformation label of the last position from the conditional probability as follows.

$$P(Y_j = y_j | X = x) = \begin{cases} \frac{1}{Z(x)} F(t, y_j, x) \exp\left(\Phi(y_j, y_{j+1})\right), j < L \\ \frac{1}{Z(x)} F(t, y_j, x), j = L \end{cases}$$

$$F(t, y, x) \begin{cases} \sum_{u=0}^{N} F(t-1, u, x) \exp(\Phi(u, y) + \Psi(y, x_i)), t > i \\ \exp(\Phi(y_{t-1}, y) + \Psi(y, x_t)), t = i, i > 1 \\ \exp(\Psi(y, x_t)), t = i, i = 1 \end{cases}$$

In the above, $Z(x)$ is the partition function and can be calculated using the forward–backward algorithm. After the conformation state at position $j$ is sampled, the algorithm replaces $j$ by $j-1$ and repeats the sampling process until position $i$ is sampled. Once the labels of the segment are sampled, we can sample the real-valued angles from the Gaussian distribution associated with a label.

*Folding simulation*: the folding simulation begins with a heating up process, in which we repeatedly sample conformations for the whole RNA using the above-mentioned segment conformation sampling algorithm. This heating up procedure terminates if one conformation without steric clashes is generated. In our experiments, we usually can obtain a conformation without clashes very quickly, which is used as the initial conformation of the simulated annealing optimization (Andrieu *et al.*, 2003; Zhao *et al.*, 2010).

To resample conformations of an RNA, we build a conformation sampling guide tree based upon the base pairing information in the RNA and all the nodes in the tree are marked as 'undone'. The torsion angles of the RNA are resampled using a bottom-up method along the tree as follows. We randomly pick up an 'undone' node $A$ in the tree, which is either a leaf node or a node with all the child nodes being marked as 'done'.

(i) If $A$ is a leaf node, we resample the angles for the segment corresponding to $A$ using the segment conformation sampling algorithm.

(ii) If $A$ has one or two child nodes, by cutting out the segments corresponding to the child nodes, we have at most three separate segments left in A, for which we use the segment conformation sampling algorithm to generate angles separately.

The new conformation is accepted if its energy is lower. Otherwise it is accepted by a probability $\exp(E/T)$, where $E$ is the energy difference between current and the new conformations and $T$ is the annealing temperature. This sampling procedure is repeated 3000 times and then node $A$ is marked as 'done'. The folding simulation process ends when the root node is marked as 'done'.

*Energy function*: different from the complex energy function in FARNA, we adopt a simple energy function used by BARNACLE (Frellsen *et al.*, 2009) as follows.

$$E = \sqrt{\frac{1}{|H|}\sum_{k=1}^{|H|}(\hat{d}_k - d_k)^2}$$

In the above formula, *H* is the number of hydrogen bonds formed in the secondary structure (every A–U and G–U pair contributes two distances, and every C–G pair contributes three distances), $\hat{d}_k$ is the distance between the donor and the acceptor of the *k*-th hydrogen bond and $d_k$ is the average length of hydrogen bonds of the same type. The smaller this value is, the more the decoy is consistent with its secondary structure. The energy is measured in Å, and the ideal base pair energy of 0 Å is only obtained for conformations with perfect base paring. We employ such a simple energy function (without any tuned parameters) so that we can carefully examine the performance of our sampling algorithm and perform a well-controlled comparison with other sampling methods such as BARNACLE. More sophisticated energy items, such as $Mg^{2+}$ ion interaction and stacking effect of base pairs, can be taken into account in future study.

### *CRF model training*

*Training data*: we build our training dataset from the RNA structure classification database DARTS (Abraham *et al.*, 2008), which collects 244 structures representing 1333 solved RNA structures and groups them into 94 clusters. Our training set comes from the 94 cluster representative structures, which have ~6000 nucleotides in total. We use all 94 cluster representative structures to build empirical distributions of bond torsion angles. To make sure our training dataset does not overlap with the 11 benchmark RNA molecules, we exclude the representative structures in the same cluster as the 11 benchmark RNA. With the remaining 83 structures, we use 3-fold cross-validation to determine the CRF model regularization factors λ and μ and the proper window size. In each fold validation, two thirds of structures are used for training and the remaining for test.

*Model selection*: the (training/test) accuracy of a second-order CRF model is defined as the number of correctly predicted states divided by the total number of positions. Fixing the number of conformation states in a CRF model, we search for the appropriate regularization factors and window size using a grid search strategy. As shown in the Supplement Figure 1 of (Wang and Xu, 2011), the CRF model with 50 conformation states has the best performance when $\lambda = 5$, $\mu = 10$ and window size = 5. We choose these parameters to maximize accuracy and avoid overfitting. A larger window size does not improve the test accuracy significantly, but increase the accuracy gap between the training and test data, which might indicate overfitting.

We also investigate the sampling performance of the CRF model with respect to the number of conformation states. We tested our CRF models with 20, 30, 50, 80 and 100 conformation states. For each CRF model, we generate 3000 decoys for each of the five RNAs: 2a43, 28sp, 2f88, 1zih and 1xjr. Figure 4-1-6 shows the 5% quantiles of the RMSD distributions for decoys generated by four different CRF models. As shown in Figure 4-1-6, the model with 50 states generates better decoys than others.

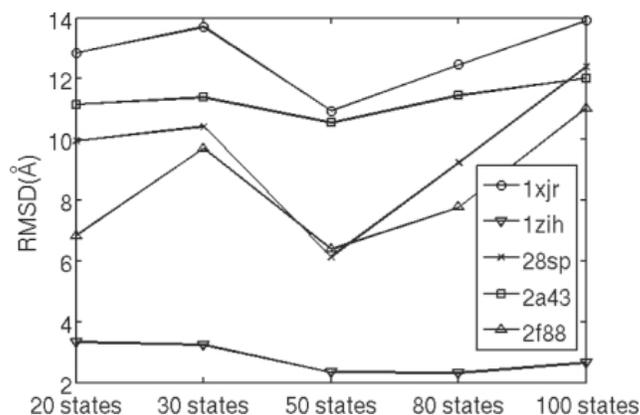

**Figure 4-1-6. The 5% quantiles of the RMSD distributions for decoys sampled from the CRF models with different number of conformation states. *Y*-axis is the RMSD value.**

Using different methods to model the distribution of torsion, $\chi$, makes a slight difference on the quality of sampled decoys. Figure 4-1-7 shows the 5% quantiles of RMSD values for 300 decoys sampled using four different χ distributions with a well-trained CRF model. In Model 1, we fix χ as the mean of the training data. Model 2 samples χ from a log normal distribution. Model 3 samples χ from a normal distribution. Model 4 uses sample χ directly from the training data without using any mathematical modeling. Finally, we decide to use the normal distribution for χ, to yield a bit of variance.

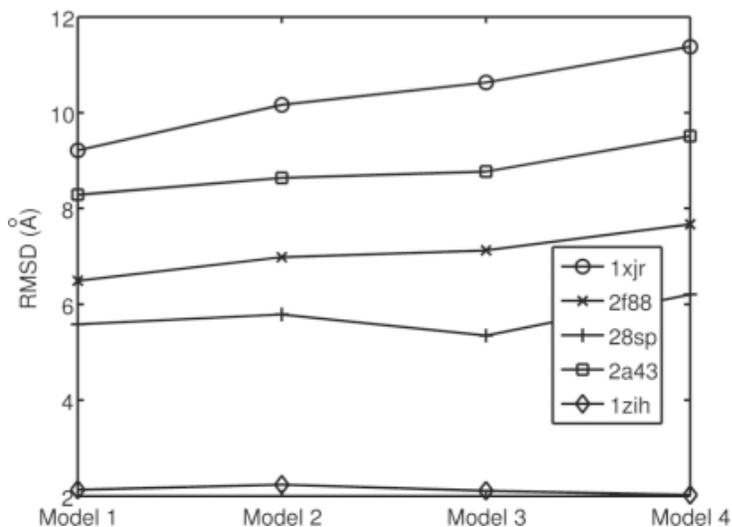

**Figure 4-1-7. The 5% quantiles of the RMSD distributions for decoys sampled from models with different distributions of torsion χ. Model 1 uses a fixed value of χ. Model 2 uses log normal distribution. Model 3 is the normal distribution. Model 4 is the empirical distribution of all values in training data. Models 2 and 3 are fit from all training data.**

**Table 4-1-1. Compared with FARNA, our method produces better decoys in terms of the best cluster centroid. The bold numbers reveals the better results. The best cluster centroid is clearly seen from the 5 clusters of all the decoys. The results of FARNA are taken from Table 1 in Das and Baker (2007). Column 'Best cluster centroid' lists the RMSD of the best cluster centroid of the top 1% decoys with**

the lowest energy. Column 'No. of decoys' is the number of decoys generated by the methods. Bold fonts indicate better results.

|  |  |  | FARNA | | | TreeFolder | | |
| --- | --- | --- | --- | --- | --- | --- | --- | --- |
| PDB ID | Method | Len | Best cluster centroid | Lowest RMSD decoy | #Decoys | Best cluster centroid | Lowest RMSD decoy | #Decoys |
| 1a4d | NMR | 41 | 6.48 | 3.43 | 28 949 | **3.65** | **2.69** | 7168 |
| 1esy | NMR | 19 | 3.98 | **1.44** | 69 103 | **2.00** | 1.52 | 22 529 |
| 1kka | NMR | 17 | 4.14 | **2.08** | 81 492 | **3.71** | 2.4 | 24 934 |
| 1l2x | X-ray | 27 | **3.88** | **3.11** | 47 958 | 8.07 | 3.97 | 15 360 |
| 1q9a | X-ray | 27 | 6.11 | **2.65** | 48 817 | **4.76** | 3.5 | 15 415 |
| 1qwa | NMR | 21 | **3.71** | **2.01** | 65 977 | 3.77 | 2.49 | 18 838 |
| 1xjr | X-ray | 46 | 9.82 | **6.25** | 24 646 | **9.26** | 7.05 | 7168 |
| 1zih | NMR | 12 | 1.71 | 1.03 | 117 104 | **1.19** | **0.73** | 40 960 |
| 28sp | NMR | 28 | 3.2 | **2.31** | 46 034 | **2.96** | **1.91** | 17 117 |
| 2a43 | X-ray | 26 | 4.93 | **2.79** | 49 972 | **4.52** | 3.47 | 18 432 |
| 2f88 | NMR | 34 | 3.63 | **2.41** | 36 664 | **3.33** | 2.7 | 12 230 |

## *Results*

We use 11 RNAs tested by both BARNACLE and FARNA to benchmark our method TreeFolder. These RNAs contain 12~ 46 nucleotides and are not homologous to any structures in our training dataset. In case an RNA has multiple NMR structures, we use the first structure in the PDB file as its native structure.

It is not very reliable to compare two methods simply using the decoys with the lowest RMSD, since they may be generated by chance and also depend on the number of decoys to be generated. The more decoys are generated, the more likely the lowest-RMSD decoy has lower RMSD from the native. Therefore, a better strategy is to compare the RMSD distributions of decoys.

*Our TreeFolder generates better decoys than FARNA*: we compare FARNA and TreeFolder in terms of the quality of the decoy clustering centroids. Similar to FARNA clustering only on the top 1% decoys with the lowest energy, we run MaxCluster to cluster the top 1% of our decoys with the lowest energy into five clusters. As shown in Table 4-1-2, TreeFolder can generate decoys with better cluster centroids for nine RNAs: 1a4d, 1esy, 1kka, 1q9a, 1xjr, 1zih, 28sp, 2a43 and 2f88. By the way, even if a significantly smaller number of decoys is generated by us, the lowest RMSD decoys by our TreeFolder for 1a4d, 1zih and 28sp still have smaller RMSD than those by FARNA.

*Our TreeFolder generates better decoys than BARNACLE*: Table 4-1-3 displays the 5% and 25% quantiles of the RMSD distributions for decoys generated by BARNACLE and TreeFolder. The quantiles by BARNACLE are taken from Supplementary Table S4 in Frellsen *et al*. (2009). BARNACLE considers only decoys with energy <1, since this kind of decoys are likely to have more correct base pairings. We use exactly the same energy function as BARNACLE, so we also consider only decoys with energy <1 to ensure a fair comparison. We did not generate as many decoys as BARNACLE and thus for some test RNAs we do not have many decoys with energy <1. In this case, we use decoys with energy <2. On the 10 RNAs shown in Table 4-1-3, TreeFolder yields better RMSD distributions for eight of them: 1esy, 1kka, 1q9a, 1qwa, 1xjr, 1zih, 28sp, 2a43 and 2f88.

**Table 4-1-3. The 5 and 25% quantiles of the RMSD distributions for decoys generated by our method TreeFolder and BARNACLE. Bold numbers indicate better distributions. Columns '#energy < 1' and '#energy < 2' list the number of decoys with energy <1 and <2, respectively. 'Bps' is the number of base pairings.**

|        |     |     | BARNACLE | | TreeFolder | | | | | |
| PDB ID | Len | Bps | 5% | 25% | 5% | 25% | # Energy <1 | 5% | 25% | # Energy <2 |
| --- | --- | --- | --- | --- | --- | --- | --- | --- | --- | --- |
| 1esy | 19 | 6 | 2.99 | 3.28 | **2.19** | **2.60** | 577 | **2.25** | **2.78** | 1102 |
| 1kka | 17 | 6 | 4.40 | 5.02 | **3.75** | **4.30** | 349 | **3.8** | **4.39** | 776 |
| 1l2x | 27 | 8 | 5.43 | 6.88 | – | – | 0 | 5.44 | 8.08 | 5 |
| 1q9a | 27 | 6 | 4.80 | 5.42 | **4.55** | **5.05** | 486 | **4.61** | **5.07** | 1025 |
| 1qwa | 21 | 8 | 4.06 | 4.64 | **3.65** | **4.26** | 407 | **3.9** | **4.51** | 884 |
| 1xjr | 46 | 15 | 10.41 | 11.01 | **8.50** | **9.43** | 22 | **8.84** | **9.79** | 540 |
| 1zih | 12 | 4 | 1.72 | 2.16 | **1.32** | **1.84** | 1721 | **1.36** | **1.88** | 1931 |
| 28sp | 28 | 8 | 3.23 | 3.76 | **2.88** | **3.43** | 152 | **2.93** | **3.58** | 563 |
| 2a43 | 26 | 7 | 4.72 | 6.08 | – | – | 0 | **4.64** | **5.48** | 26 |
| 2f88 | 34 | 13 | 3.82 | 4.41 | **3.73** | **3.73** | 1 | 3.85 | 4.57 | 130 |

*Sequence information is important for RNA conformation sampling*: different from other two state-of-art methods, FARNA and BARNACLE, our TreeFolder makes use of sequence information to significantly improve conformation sampling, as measured by the median RMSD values of decoys. The result is shown in Table 4-1-4, in which we compare two CRF models: one using sequence to sample conformations and the other not. Without using sequence information, our CRF method is similar to BARNACLE. That is, it models only angle state transitions in a RNA structure. Both CRF models use 50 conformation states. For the CRF model without sequence features, the regularization factor is set to 5 (i.e. $\lambda = 5$). While for the CRF model utilizing sequence information, the regularization factor are set to 5 and 10 (i.e. $\lambda = 5, \mu = 10$). To calculate the median RMSD, for each RNA we generate 300 decoys using the two CRF models.

Table 4-1-4. Comparison between the CRF models using or without using sequence information. For 10 of the 11 tested RNAs, the model using sequence information yields decoys with much smaller median RMSD. Bold numbers indicate smaller RMSD values.

| | Median RMSD value | | | Median RMSD value | |
| --- | --- | --- | --- | --- | --- |
| PDB ID | With seq. feature | Without seq. feature | PDB ID | With seq. feature | Without seq. feature |
| 1zih | **2.68** | 4.56 | 28sp | **6.02** | 10.27 |
| 1esy | **3.73** | 6.17 | 1a4d | **7.79** | 11.60 |
| 1kka | **5.49** | 6.67 | 2a43 | **10.62** | 12.25 |
| 1qwa | **5.58** | 5.99 | 1l2x | 11.01 | 10.74 |
| 1q9a | **5.91** | 6.84 | 1xjr | **10.92** | 12.70 |
| 2f88 | **6.36** | 9.55 | | | |

*Sampling real-valued angles generates better decoys*: in order to show the detailed difference between our TreeFolder and FARNA, we look into the decoys of 1esy. We choose it because that FARNA and TreeFolder yield the largest difference on this RNA among all the 11 tested RNA molecules. As shown in Figure 4-1-8. TreeFolder can generate a much larger percentage of decoys with RMSD < 4 Å than FARNA. We also compute local RMSD of each position in the decoys, which is defined as the RMSD of the segment of four consecutive nucleotides starting with this position, as compared to the native structure. We calculate the correlation between the local RMSD of each position with the global RMSD, as shown in Figure 4-1-9. Among the decoys generated by both FARNA and TreeFolder, the local RMSD at position 13 has the highest correlation with the global RMSD. We also calculate the angle error at each position by Error = $\|v - v_0\|_2$, where $v$ is the angle vector of a decoy at one position and $v_0$ is the native angle vector at the same position.

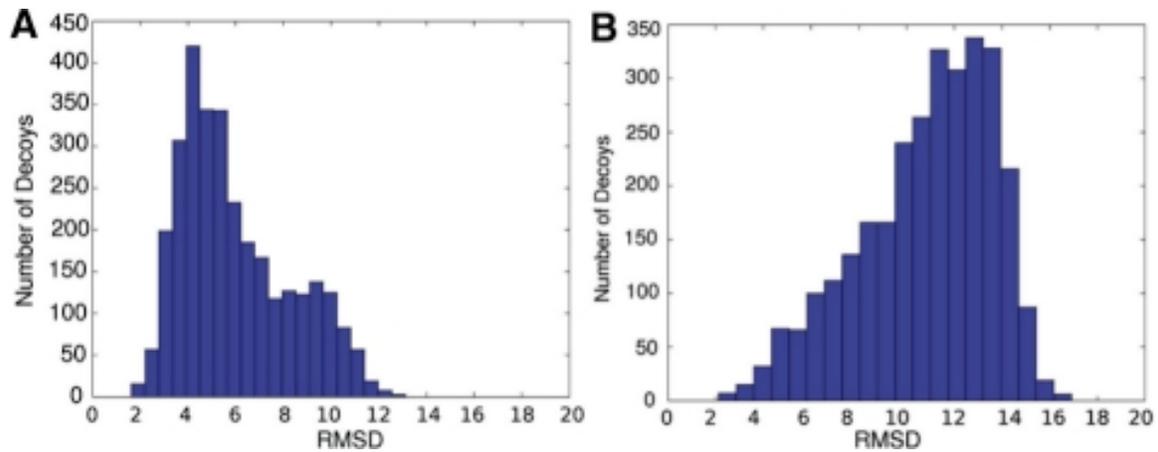

Figure 4-1-8. The RMSD histograms of the 3000 decoys generated by our method TreeFolder (A) and FARNA (B) for 1esy.

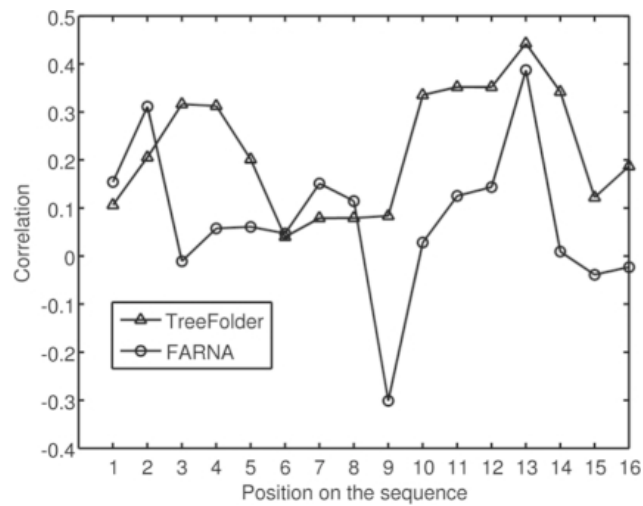

Figure 4-1-9. Correlation between the local RMSD at each position and the global RMSD. The *X*-axis is the start position of a segment.

Figure 4-1-10 shows the angle error histograms in three positions 13, 14 and 15. The angles at these three positions determine the conformation of the segment starting at position 13. At positions 13 and 15, the angle errors by our method TreeFolder are significantly smaller than those by FARNA. As Figure 4-1-10 shows, the angle errors by FARNA are distributed around several separated peaks, which may be caused by the limited number of fragments used in FARNA. In contrast, the angle errors by TreeFolder are distributed more smoothly, possibly because we can sample real-valued angles.

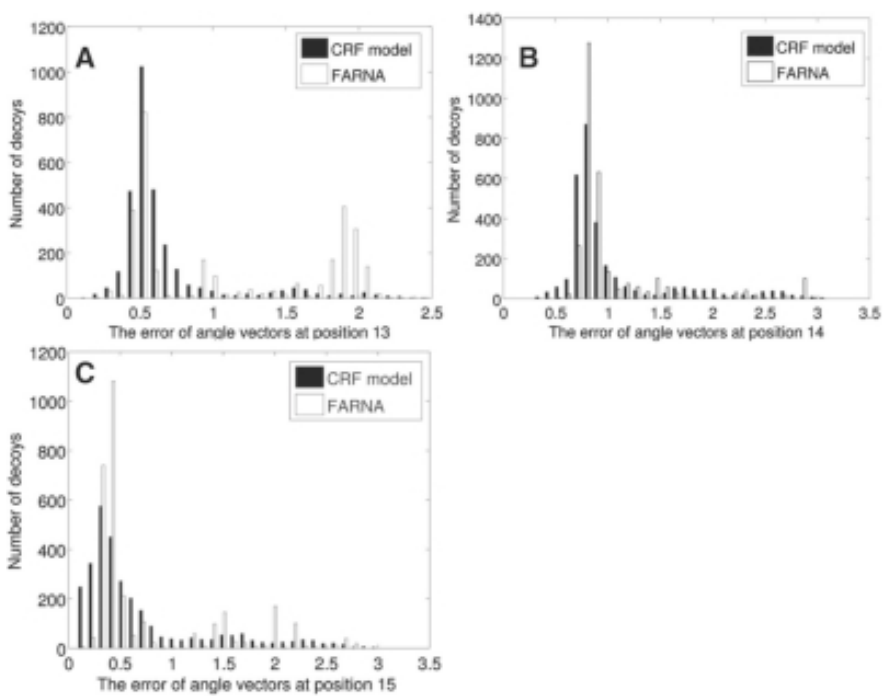

**Figure 4-1-10. The angle error histograms at positions 13, 14 and 15. At positions 13 and 15, the decoys by our TreeFolder have much smaller angle errors than those by FARNA.**

*Folding RNA using predicted secondary structures*: we use the secondary structures predicted by CONTRAfold (Do *et al.*, 2006) and sample 1000 decoys for each RNA. The quantiles of their RMSD values are shown in Table 4-1-5. On 6 of the 10 tested RNA, decoys generated from native secondary structures are better than those from predicted secondary structures. On the other four RNAs, the difference between the two types of decoys is small, because of accurate secondary structure prediction. The results for 1l2x and 2a43 from predicted secondary structures are quite bad, since all of their base pairs are contained in a H-type pseudoknot and only half of their base pairs are recovered by CONTRAfold. However, our TreeFolder generates decent conformations for half of the pseudoknot with predicted base pairs, as shown in brackets. In particular, TreeFolder generates decent structures for 2a43 from nucleotides 1 to 14 and for 1l2x from nucleotides 1 to 18, respectively. In order to improve sampling performance on the whole structures of 1l2x and 2a43, we need an energy function like what is used in FARNA to guide the folding simulation.

**Table 4-1-5. Comparison between folding with native and predicted secondary structure. The four numerical columns list the RMSD values of the 5th and 25th quantiles of the decoys with energy values <2. Bold numbers indicate better results.**

| PDB ID | Distribution of RMSD values | | | |
|---|---|---|---|---|
| | Native SS | | Predicted SS | |
| | 5% | 25% | 5% | 25% |
| 1esy | **2.25** | **2.78** | 3.90 | 4.35 |
| 1kka | **3.80** | **4.39** | 4.57 | 5.46 |
| 1l2x | **5.44** | **8.08** | 15.23 (3.53) | 17.32 (3.88) |
| 1q9a | 4.61 | 5.07 | 4.65 | 5.01 |
| 1qwa | 3.90 | 4.51 | 3.45 | 4.31 |
| 1xjr | **8.84** | **9.79** | 9.17 | 9.79 |
| 1zih | **1.36** | **1.88** | 3.56 | 4.02 |
| 28sp | 2.93 | 3.58 | 2.71 | 3.63 |
| 2a43 | **4.64** | **5.48** | 21.22 (3.89) | 21.99 (4.35) |
| 2f88 | 3.85 | 4.57 | 3.58 | 4.21 |

*Comparison with MC-Sym on the large RNA molecules*: our TreeFolder is much faster than the MC-Fold and MC-Sym pipeline (Parisien and Major, 2008) for folding large RNA molecules, as shown in Table 4-1-6. The running times in this table were obtained on a workstation with 96 GB RAM and 24 computing cores [2.67 GHz Intel(R) Xeon(R)].

**Table 4-1-6. Running time comparison between MC-Sym and our TreeFolder on large RNA molecules.**

| PDB ID | Length | MC-Sym (h) | TreeFolder (s) |
| --- | --- | --- | --- |
| 1l8v | 152 | 48 | 1919 |
| 2gis | 94 | 32 | 564 |
| 1vc7 | 74 | 46 | 400 |

*Overlay examples*: Figure 4-1-11 shows three overlay examples of 1q9a, 2a43 and 1xjr with length of 27 nt, 26 nt and 49 nt, respectively. Pictures in blue display native, while in red the best centroids produced by our algorithm. As shown in this figure, our algorithm recovered a pseudoknot for 2a43.

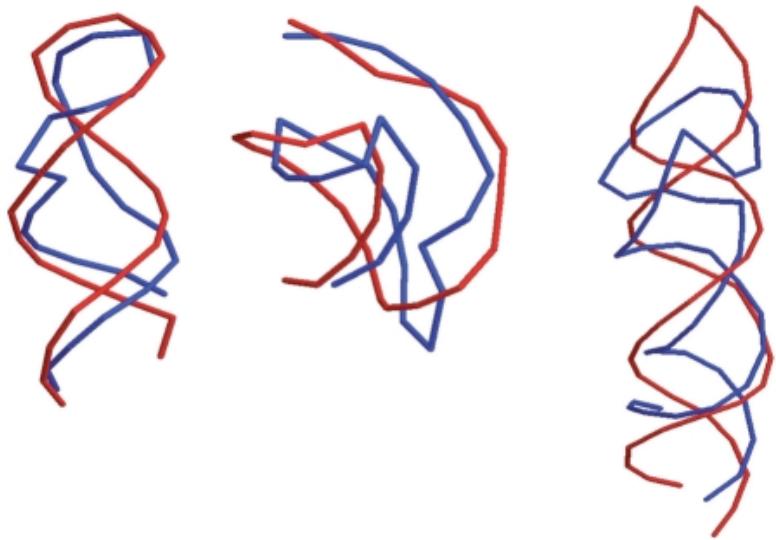

Figure 4-1-11. Overlay representation of the best centroids (red) of 1q9a, 2a43 and 1xjr (from left to right) with their native structures (blue). These three RNA molecules have lengths of 27 nt, 26 nt and 49 nt.

***Conclusions***

We have presented a new method TreeFolder for modeling RNA sequence–structure relationship and conformation sampling using CRFs and a tree-guided sampling scheme. Our CRF method not only captures the relationship between sequence and angles, but also models the interdependency among the angles of three adjacent nucleotides. Our conformation sampling method distinguishes from FARNA in that we do not use fragments to build RNA conformations, so that we do not need to worry about if there are a sufficient number of structure fragments to cover all the possible local conformations. Our TreeFolder also differs from both FARNA and BARNACLE, in that we use primary sequence to estimate the probability of backbone angles, while the latter two do not. In addition, we also use a tree, built from (predicted) secondary structure, to guide conformation sampling so that at one moment we can simultaneously sample conformations for two segments far away from each other along the RNA sequence. In contrast, both FARNA and BARNACLE can only sample conformations for a single short segment at any time. The results indicate that our

TreeFolder indeed models sequence–structure relationship well and compares favorably to both FARNA and BARNACLE, even if we use only the same simple energy function as BARNACLE.

We will extend our TreeFolder further. For example, we can incorporate information in sequence homologs into our CRF model so that we can estimate the conformation probability more accurately and thus improve the sampling accuracy. Information in homologs has been successfully used in RNA secondary structure and should be useful for 3D structure prediction. Information in homologs has also been used for protein conformation sampling (Zhao *et al.*, 2010). Currently TreeFolder works well when the native base pairing information is used to calculate the energy function (same as BARNACLE) and to build the sampling guide tree. Not all the RNAs without 3D structures have the native base pairing information. Our next step is to further improve TreeFolder with the predicted base pairings. In particular, we need to design an energy function similar to what is used in FARNA to guide the folding simulation so that TreeFolder works well even if the predicted secondary structure is not very accurate. To tolerate errors in the predicted base pairing information, we will use the predicted confidence as the weight of each item in the energy function and only use those base pairings with high confidence to build the conformation sampling guide tree. We can also take another strategy to circumvent possible impact of errors in the predicted base pairings. In particular, we will extend our CRF method so that we can simultaneously sample base pairings and 3D conformations so that errors in the predicted base pairings will be corrected in the folding simulation process.

Currently, we use a very simple energy function to guide the folding simulation. We will develop a more sophisticated energy function to guide the formation of hydrogen bonds in a better way, just like what FARNA does. Thus, we cannot only generate decoys with better RMSD, but also with better hydrogen bonds. Our TreeFold algorithm shows that using the intermediate information, even if predicted, will help improve 3D structure prediction.

## 4.2 Intermediate states used for protein 3D structure prediction

Our previous studies illustrate how to define the macromolecular intermediate states, and show the results we have achieved on intermediate states prediction. Based on these results, we propose a novel method for protein 3D structure prediction by modeling the intermediate states. In this study, we formulate a new protein 3D structure potential involving a machine learning model between sequence information and intermediate state. The novel statistical energy function is consist of a neural network taking input of features used in protein secondary structure prediction and contact map prediction, including PSSM and mutual information. The program of protein secondary structure and contact map prediction will produce the probability results for each amino acid and each amino acid pair in the sequence. The potential energy function will takes the probability results as input to predict the distance for each pair of the amino acids. With the pair distance probability, the potential energy function is defined as a likelihood of a given protein 3D structure candidate, a decoy.

Before we introduce our method, we want to briefly introduce the background of how to use energy function to do protein 3D structure prediction. Energy function, also named potential energy function, evaluates the quality of protein 3D structure. Potential energy is derived from biophysics, where the potential is defined as the energy used to decompose ensemble of atoms. Physical potential of a protein crystallized 3D structure can be calculated as the total energy inside all chemical bonds between each pair of atoms. According to physical theory, the crystalized structure of a protein molecule is most stable status, which corresponding to the lowest potential energy. In molecular dynamics, the physical potential energy is computed as a simulated thermodynamics process moving the positions of all the atoms. However, the computation of physical potential is not feasible due to computation resource limit and may not be necessary when considering the trade-off between computing complexity and the precision of protein 3D structure prediction.

The statistical potential energy is an approximation to the physical potential to evaluate the stability of a given protein 3D structure. Different from physical potential energy, the statistical potential energy is optimized from the observation of known 3D structures. [cite dope]. Compared with physical potential energy, statistical energy function is simpler in computation, and can be defined a high level of resolution, i.e. amino acids instead of atom level. Many studies have been done on the statistical potential (Shen 2006).

### *Method*

Our new statistical energy function establishes a non-linear relationship between the features of the protein sequence and the distance of all atom pairs by using a neural network model. In our model, the responsible variable is the discretized distance between each pair of atoms, which ranges from 0 to 11, corresponding to 12 distance intervals. Our features from the protein sequence include the position specific scoring matrix, which is the mutation rate for each position, and represents short-range sequence feature. The global sequence features are calculated from the co-evolutionary relationship between each pair of amino acids, which are mutual information and decoupled mutual information.

The neural network used in our potential energy function predicts the categorical distance label from the given protein amino acid sequence and its feature vectors. The output of the neural network is probability for each label category. This is done by a layer of restricted Boltzmann machine (Salakhutdinov and Hinton, 2009). Thus, the general form of our statistical potential energy can be written as the following formula.

$$P(Y = y | X = x) = \exp\left(-w_{y,k} NN_k(x)\right) / Z(x) \quad (4\text{-}2\text{-}1)$$

In the above, $NN_k$ is the output value of the $k$-th on the last layer. $\{w_{y,k}\}$ is a 12x40 matrix for the weight of linear combination of the Boltzmann machine layer. $Z$ is the normalization factor.

With the probability of the distance of each atom pair, we can define the energy function of a given 3D structure of protein. The whole structure is decomposed into the pairwise distance $d_{ij}$ of all the beta atom pairs on the given protein sequence. The statistical potential energy function $E$ of the given whole structure is defined by the following formula.

$$E = \sum_{i,j} E(d_{i,j}, x_{i,j}) = -\log[P(Y = d_{i,j} | X = x_{i,j}) / R(i, j, d_{i,j}, L)]$$

In the energy function of the whole structure, the $d_{i,j}$ is the discretized distance between the atom $i$ and $j$. $x_{i,j}$ is the feature of the atom pair $(i,j)$, and $R(i, j, d_{i,j}, L)$ is the reference factor of the pairwise distance, defined as following formula.

$$R(i,j,d_{i,j},L) = \frac{\text{Number of atom pairs with distance } d_{i,j}}{\text{Total number of atom pairs in the protein with length of } L}$$

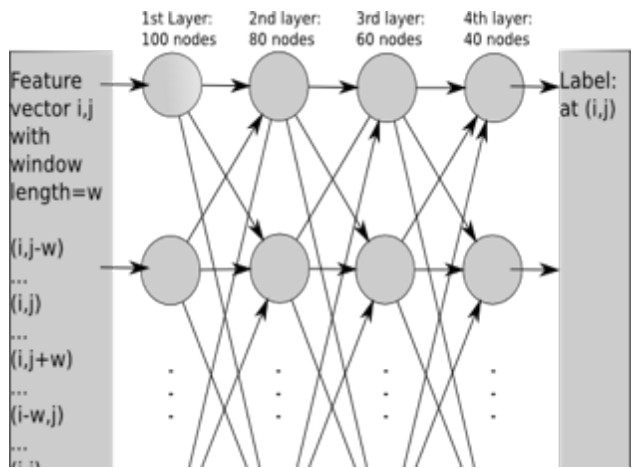

**Figure 4-2-1. The neural network model used in our potential energy function.**

*Feature vector*

The feature vector used in our model is built as the following procedure. First, we build the multiple sequence alignment of the given protein sequence. With the multiple sequence alignment, we compute most important features we used in our energy function. The evaluation of a decoy with our energy function requires a reference factor for each pair of amino acids. In other words the potential energy of a given decoy is decided by both the energy function valued computed from each pair of amino acids and the reference states for the pairs.

The feature vector for each pair of (i,j) is consist of the position specific scoring matrix (PSSM), which is a 20 vector for each position. The PSSM matrix is resulted from homologous sequence search in a non-redundant sequence database by NCBI PSI-BLAST. This PSSM feature is proved very helpful in many sequence related research (Jones, 1999; Källberg, et al., 2012; Ma, et al., 2014; Peng and Xu, 2010; Peng and Xu, 2011; Söding, 2005; Wang and Xu, 2013; Wang, et al., 2011). We also include the generalized pairwise mutual information feature set. This feature set includes the mutual information, defined as following.

$$Mu(i,j) = \sum_{a,b} \log \frac{f_{i,j}(a,b)}{f_i(a)f_j(b)}$$

In this formula, $f_{i,j}$ is the frequency of two amino acid at the position (i,j).

*Neural Network Optimization and Model Selection*

Our energy function contains a 4-layer neural network. For every layer in the neural network, there are 100, 80, 60, and 40 nodes responsively. The first layer with 100 nodes takes input from features, and the last layer output will output the probability values of 13 states with soft-max as shown in Equation (4-2-1). The nodes between each two neighbored layers are fully connected, e.g. for each node in the second layer or layers after, it takes input from all the output of its previous layer.

There are an enormous number of the atom pairs, which have the longest distance label. The pair with this label has a distance larger than 15 angstrom (A). Including all of them in the training process not only slows down the optimization, but not benefit the accuracy very much. We have watched the optimization process with 30% and 10% of all the pairs with larger than 15A distance. The decreasing of the object function on a validation dataset is much similar from two difference optimization processes. So we only include 10% of the pairs with the last state in our training set.

The weights of our neural network model show the importance of the co-evolutionary information we used. In the following figure, we estimate the importance for each pair of feature and label by using the product of all the edge weight from the feature to the label. The importance for all 13 states can be shown on the following figure. We can see the signals on the feature from 1360 to 1575 show stronger the importance score, which are associated with the co-evolution features in our new method.

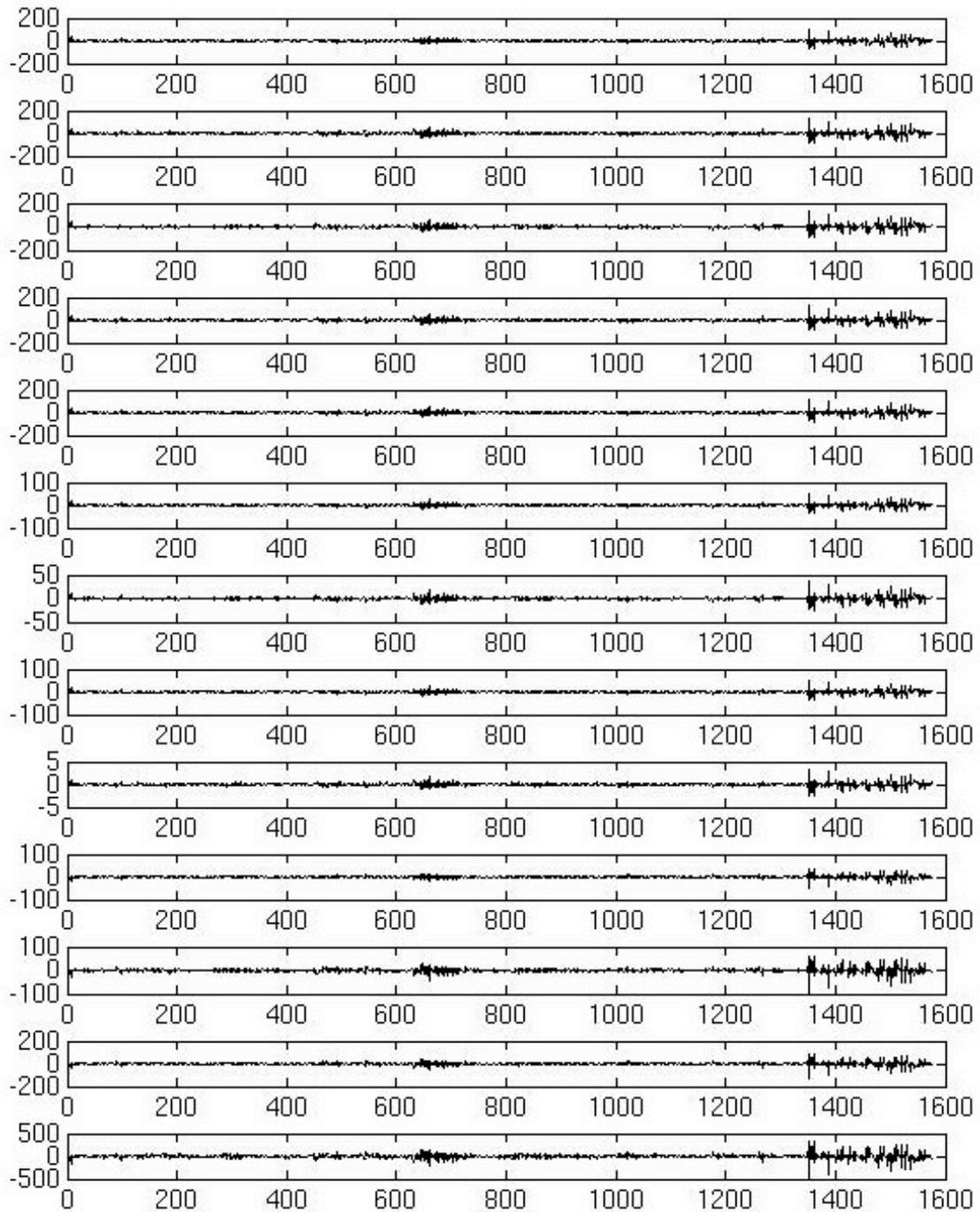

**Figure 4-2-2.** The importance score for each pair of feature and label. Each label is corresponding to a subplot from the top to the bottom. X axis is the importance score for each feature and Y axis is the index of the feature.

From the network model weights, we also find that it is not necessary to divide the distance larger than 16A. We compute the correlation between the weights used in the soft-max layer among the 13 states and show them in the Figure 4-2-2. The last 4 states have highly correlated weights in their soft-max layer, which implies the model has difficult to distinguish them.

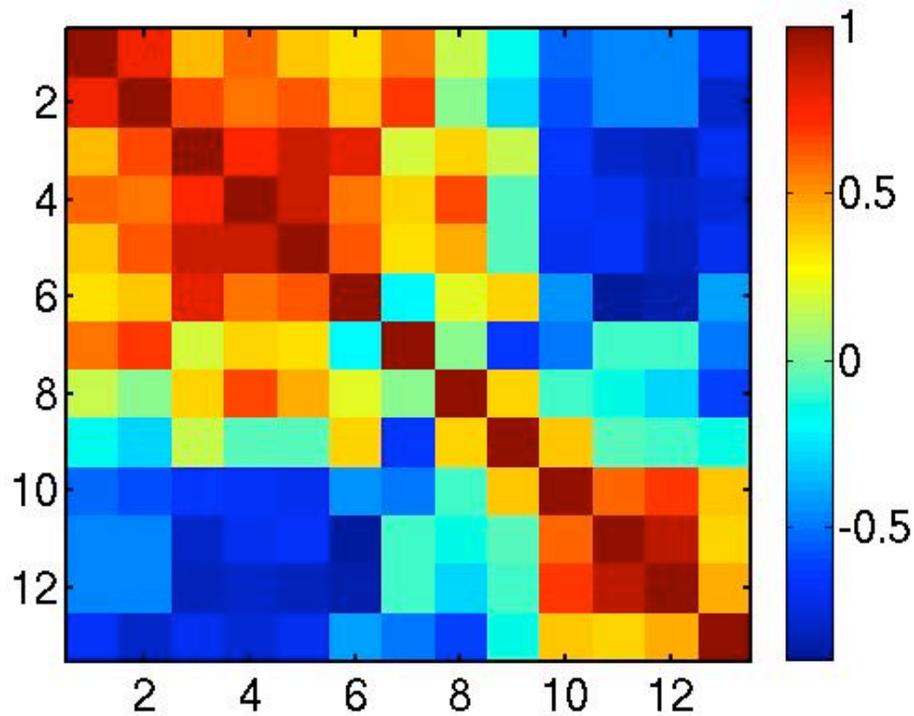

**Figure 4-2-3. The correlation among the weights of the soft-max layers for the 13 states.**

*Training Dataset*

The training set is built based on 1200 non-redundant protein 3D structures. The redundancy between each pair of proteins in our dataset is smaller than 25% sequence identity. Since the number of amino acid pairs is proportional to the square of the length of the protein sequence, we limit the proteins with length up to 350 amino acids in our training dataset.

*Data preprocessing*

We remove the sequence that has missing amino acids in test datasets and training datasets. We also remove sequence shorter than 50 amino acids, which usually has few distance contact in its 3D structure.

*Computing environment*

Our training program takes 300 cores and 72 hours to finding a local optimal solution for the weights for the neural network. We used L-BFGS algorithm to optimize the weight of the neural network (Liu and Nocedal, 1989).

*Results*

We first compare our new energy function with state-of-the-arts methods on decoy discrimination. We measure the improvement of decoy discrimination by 5 metrics, number of correctly identified natives, ranks of the native structure, GDT score of the first-ranked decoy, and correlation of the energy and the decoy quality, and Z score of the native energy.

For each target there are dozen to hundreds decoys in the dataset of this experiments. The number of correctly identified natives is the number of proteins which native structures are ranked lowest among all the decoys. The rank of the native structure shows the rank of the native structure of a protein among all the decoys in the order of potential energy values. GDT score is used to compare the lowest energy decoy and its native structure. The lower the GDT score is the better the quality of the decoy. We also calculated the Pearson correlation coefficient between the potential energy value and the GDT score. A good energy function should produce small scores for high quality decoys and high scores for low quality decoys. Numerically, the quality of an energy function is showed by the correlation between decoy energy values and decoy GDT scores. We also calculate the "best GDT score", which is the GDT score for the decoy with lowest energy value with a given energy function. "Native rank" is the place where the protein native structure lie in the order of energy values.

On the Rosetta dataset (Raman, et al., 2009), we compare our method with other state-of-the-arts methods, including DOPE, DFIRE, MyDope and OPUS. We first examine our method on ranking the decoys. In the dataset of Rosetta, there are 58 proteins and 120 decoys for each protein. The decoys are generated by Rosetta protein structure sampling method.

**Table 4-2-1. Performance on Rosetta dataset. We also compare our the new method with the Epad method on three datasets, the Rosetta decoy set, I-Tasser decoy set (Wu, et al., 2007) and the CASP5-8 targets with an extensive set of metrics.**

| Rosetta | DOPE | DFIRE | MyDope | OPUS | Epad | Epmi |
|---|---|---|---|---|---|---|
| Native identified | 11 | 12 | 10 | 6 | 17 | 19 |
| mean of correlation | -0.24 | -0.20 | -0.21 | -0.15 | -0.44 | -0.49 |
| z-score of native | -1.51 | -0.66 | -1.23 | 0.25 | -1.02 | -1.23 |
| mean of best GDT | 0.47 | 0.48 | 0.48 | 0.46 | 0.64 | 0.68 |
| Average native rank | 18.7 | 30.7 | 21.7 | 55.3 | 31.9 | 29.7 |

**Table 2-2-2. An extensive comparison of our new method and Epad energy function on 3 datasets, rosetta, casp5-8, itasser. Dfire, opus, rw, epad, and epmi are the energy functions for each column result.**

|  |  | rosetta | | | | | casp5-8 | | | | | itasser | | | | |
|---|---|---|---|---|---|---|---|---|---|---|---|---|---|---|---|---|
|  |  | dfire | opus | rw | epad | epmi | dfire | opus | rw | epad | epmi | dfire | opus | rw | epad | epmi |
| native-free | correlation between GDT and energy value | -0.44 | -0.35 | -0.46 | -0.46 | -0.55 | -0.58 | -0.49 | -0.65 | -0.71 | -0.84 | -0.44 | -0.29 | -0.49 | -0.52 | -0.61 |
| native-free | correlation between TMscore and energy value | -0.43 | -0.34 | -0.44 | -0.43 | -0.53 | -0.56 | -0.45 | -0.63 | -0.72 | -0.84 | -0.43 | -0.29 | -0.48 | -0.52 | -0.60 |
| native-free | GDT score of the top ranked decoy | 0.53 | 0.53 | 0.53 | 0.51 | 0.55 | 0.67 | 0.63 | 0.69 | 0.66 | 0.74 | 0.60 | 0.57 | 0.60 | 0.58 | 0.61 |
| native-free | place(rank) of the lowest energy decoy in the order of GDT | 35.76 | 36.41 | 37.33 | 38.25 | 26.90 | 4.85 | 6.33 | 4.22 | 5.30 | 2.84 | 115.09 | 175.68 | 126.11 | 163.79 | 99.43 |
| native-free | TMscore of the top ranked decoy | 0.50 | 0.50 | 0.50 | 0.49 | 0.56 | 0.75 | 0.71 | 0.77 | 0.75 | 0.82 | 0.59 | 0.55 | 0.57 | 0.56 | 0.59 |
| native-included | GDT score of the top ranked decoy | 0.70 | 0.86 | 0.71 | 0.64 | 0.86 | 0.94 | 0.86 | 0.92 | 0.70 | 0.84 | 0.93 | 0.96 | 0.98 | 0.68 | 0.95 |
| native-included | number of native structure identified in the top 5 models | 33 | 41 | 30 | 18 | 33 | 126 | 90 | 121 | 80 | 107 | 48 | 50 | 55 | 22 | 50 |
| native-included | place(rank) of the lowest energy decoy in the order of GDT | 28.90 | 14.12 | 28.45 | 34.12 | 11.26 | 0.89 | 2.45 | 1.14 | 4.98 | 1.71 | 19.77 | 24.93 | 4.61 | 139.89 | 10.05 |
| native-included | rank of the native structure | 19.57 | 10.82 | 22.61 | 32.49 | 17.12 | 0.38 | 5.13 | 0.74 | 3.89 | 2.46 | 24.50 | 26.71 | 0.80 | 77.04 | 14.73 |
| native-included | TMscore of the top ranked decoy | 0.68 | 0.85 | 0.68 | 0.63 | 0.85 | 0.95 | 0.89 | 0.94 | 0.78 | 0.90 | 0.92 | 0.95 | 0.98 | 0.66 | 0.94 |
| native-included | Z-score of the native structure energy value | -1.73 | -2.74 | -1.66 | -1.02 | -2.39 | -1.55 | -0.68 | -1.50 | -0.74 | -1.02 | -3.27 | -5.03 | -4.46 | -1.17 | -2.91 |

The other important usage of statistical energy function is in the for protein structures (Wang and Xu, 2011; Zhao, et al., 2008). We compared our novel energy function and the energy function derived without using the direct information on the dataset of CASP10 free modeling targets. The dataset we used is the CASP10 free modeling targets less the domains Quark or Rosetta Server have no submission and the long targets without domain cutting. Our method tops on 4 target domains out of 12 domains, and on 7 of 12 domains our method is better than Quark and Rosetta.

**Table 4-2-3. The sampling results on CASP 10, 12 free modelling domains.**

| | Our method outperforms the top model |
|---|---|
| | Our method outperforms RaptorX-Roll |
| | Our method outperforms Quark and Rosetta server |

| | Our new method | Quark | Rosetta server | RaptorX-Roll | The top model submitted |
|---|---|---|---|---|---|
| T0653-D1 | 0.1861 | 0.4181 | 0.4368 | 0.1656 | 0.4280 |
| T0658-D1 | 0.2229 | 0.2970 | 0.1920 | 0.2276 | 0.2910 |
| T0666-D1 | 0.4409 | 0.2252 | 0.2499 | 0.2642 | 0.4160 |
| T0684-D2 | 0.2652 | 0.2448 | 0.2711 | 0.2287 | 0.2820 |
| T0693-D1 | 0.2527 | 0.3287 | 0.2410 | 0.2672 | 0.3460 |
| T0719-D6 | 0.2366 | 0.2331 | 0.2220 | N/A | 0.3230 |
| T0726-D3 | 0.3000 | 0.1804 | 0.2331 | N/A | 0.2570 |
| T0734-D1 | 0.2975 | 0.2155 | 0.2446 | N/A | 0.2670 |
| T0735-D2 | 0.4264 | 0.3549 | 0.3504 | 0.3971 | 0.3970 |
| T0737-D1 | 0.2871 | 0.3175 | 0.3496 | 0.2816 | 0.3630 |
| T0740-D1 | 0.3110 | 0.2678 | 0.2689 | 0.4778 | 0.3610 |
| T0741-D1 | 0.1992 | 0.1394 | 0.1752 | 0.1627 | 0.2020 |

The advantage of our method is also verified by the difference between the targets our method achieve TMscore > 0.4. On these 5 targets, we run the Rosetta server, which produced results with TMscore < 0.4 on 4 of the 5 targets.

**Table 4-2-4. The targets on which our method results >0.4 in TMscore.**

|  | Our method | Robetta |
|---|---|---|
| T0651-D1 | 0.4272 | 0.3336 |
| T0651-D2 | 0.4527 | 0.2724 |
| T0663-D1 | 0.4422 | 0.4312 |
| T0663-D2 | 0.5781 | 0.7289 |
| T0726-D2 | 0.5922 | 0.3160 |

***The folding result for CASP 10 targets***

We test our method on the human-server CASP10 targets. Among 67 targets, we find our method improved the sampling decoy quality on 6 targets as shown in the following figure.

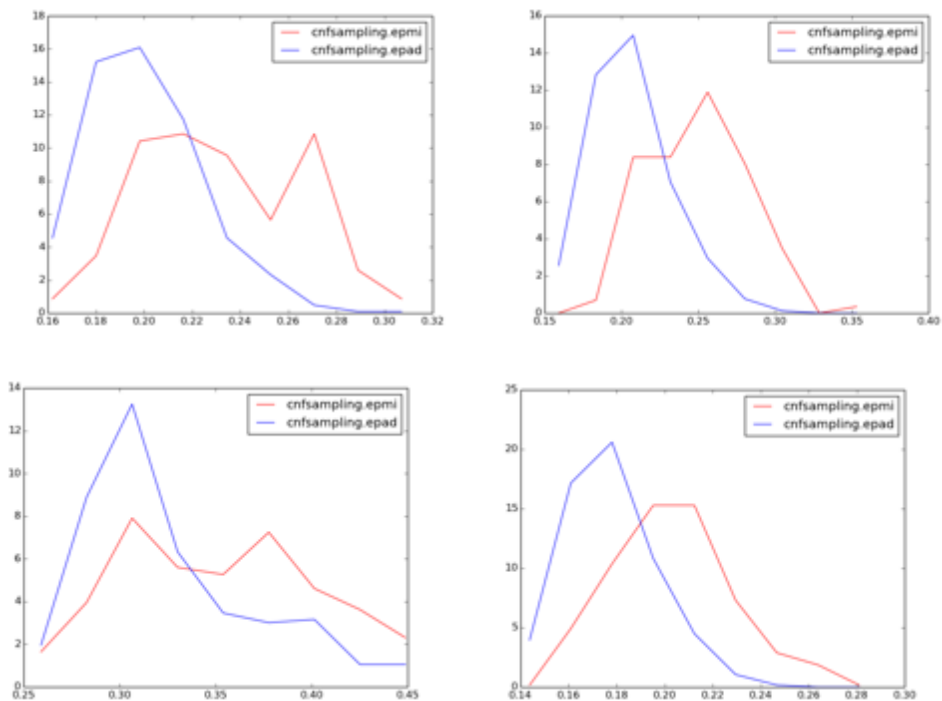

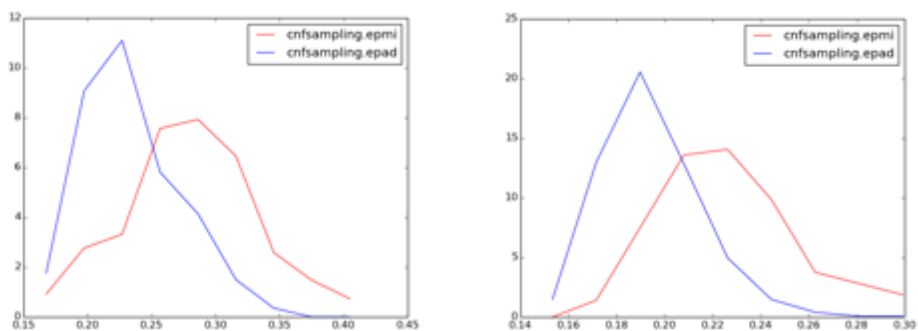

**Figure 4-2-4. Six CASP10 human-server targets (from up left to down right, T0713-D1, T0713-D2, T0651-D1, T0690-D1, T0690-D2, and T0707-D1) have improved decoys in the 3D structure sampling guided by our new energy function than the EPAD energy function.**

## *Conclusions*

In this study, we have proposed a novel potential energy function based on the pairwise co-evolutionary information. Our numerical experiments on different datasets showed its improvement on both ranking decoys and 3D structure sampling. Our statistical energy function models the atom pairs independently, i.e. we assume the probability of pair amino acid distance is independent from each other. This assumption is first used in the definition of mean field model (Le Boudec, et al., 2007), and make the computation possible on a large dataset. However, the distance of all amino acid pairs may not be independent in the complicated protein structure problem. Therefore we design a reference factor for each pair to correct the total energy to make up the dependency.

There are several ways to improve on the statistical potential function in the future based on our results achieved in this study. It will be helpful to optimize the discretized label for the distance of each atom pair. Figure 4-2-3 suggests a correlation between several discretized labels. To make use of this correlation, we can group some neighbored labels together, which would increase the sample size for each label. The reference factor is not fully developed by far. It is importance to combine the independent pair probability into the evaluation function of the whole structure. It would be helpful to model the reference factor with global constraints. Previous we have demonstrated how the global constraints can be used to improve the contact map prediction. Thus, it implies us to combine the co-evolutionary information and global constraints on the potential energy function in the future.

## Chapter 5. Summary and Future Works

Protein and RNA 3D structure prediction is a very challenging problem in the family of machine learning applications. Its difficulties include that it is a structural learning problem, its objective function is non-convex, and the dataset of training samples is still sparse and expensive to get. As a structural learning problem, the macromolecular 3D structure has many constraints, such as physical clash constraints, which disallow different atoms occupy the same position in the space, and the electrostatic preference between atoms. The objective function of 3D structure is a metric between a decoy and its native structure. Most metrics, including RMSD, GDT-score and TM-score, are not convex, which bring difficulty to design an algorithm to use the final metric as an optimization target.

Another challenging side of the problem is the evolving of homologous information, which is proved as an important feature in many sequence related problems. The homologous matrix is highly dependent on the database of all the sequences people have discovered, which is changing with the advance of technology. This makes the input feature may vary if a new database is used in building the feature. Changing the parameter of the algorithm of the multiple sequence alignment also results in a different feature. There are few works to investigate the relationship between homologous search algorithm and the performance of structure prediction.

The most important contribution of this thesis is the novel concept of intermediate state and its application in 3D structure prediction. In the native structures, we can observe many patterns and pattern combinations. These patterns are supposedly a higher level of intermediate states than secondary structure and contact map. However, current 3D structure prediction methods hardly model or search in a space with these patterns. One reason may be these patterns are often with many amino acids or nucleotides far from others on the sequence. The other reason is we still do not understand how these patterns are formed during the real folding process. In this thesis, we propose a method to make our preliminary attempt to connect intermediate states prediction and the 3D structure prediction a together by making using the observed patterns.

Through a series of research, we have proved the improvement of integrating different levels of information together. From the secondary structure prediction, we have proved the effectiveness of using the homologous information. From using the secondary structure prediction results in protein contact map prediction, we showed a method to predict the intermediate states. RNA sampling study showed a method to make use of secondary structure prediction results for RNA 3D structure prediction. Finally, we have showed that the protein intermediate state, including secondary structure and contact map, can be used for protein 3D structure prediction.

Our work in this thesis is to build a chain of machine learning models to improve macromolecular 3D structure prediction. Further improvement around the corner includes using deep learning method to automatically discover the intermediate state and optimizing each model on the logic chain. With the concept of intermediate state, we may understand the macromolecular 3D structure better.

# Reference

Abraham,M. *et al,* Analysis and classification of RNA tertiary structures. *RNA* 2008, **14**, 2274–2289.

Akutsu, T. Dynamic programming algorithms for RNA secondary structure prediction with pseudoknots. *Discrete Applied Mathematics* 2000;104(1):45-62.

Alkan,C. *et al.* RNA secondary structure prediction via energy density minimization. *Res. Comput. Mol. Biol.* 2006, **3909**, 130–142.

Asai K, Hayamizu S, Handa KI. Prediction of protein secondary structure by the hidden Markov model. Bioinformatics. 1993; 9:141–146.

Aydin Z, Altunbasak Y, Borodovsky M. Protein secondary structure prediction for a single sequence using hidden semi-Markov models. BMC Bioinformatics. 2006; 7:178.

Badorrek, C.S. *et al.* Structure of an RNA switch that enforces stringent retroviral genomic RNA dimerization. *Proc. Natl Acad. Sci.*, 2006, **103**, 13640–13645.

Backofen, R. *et al.* RNAs everywhere: genome-wide annotation of structured RNAs. J. Exp. Zool. Part B: Mol. Dev. Evol. 2007, **308B**, 1–25.

Backofen,R. *et al.* Sparse RNA folding: time and space efficient algorithms. Com.l Pattern Matching 2009, **5577**, 249–262.

Bau ,D. et al. Distill: a suite of web servers for the prediction of one-, two-and three-dimensional structural features of proteins. BMC Bioinformatics 2006, 7, 402.

Berman,H.M. *et al.* The nucleic acid database. A comprehensive relational database of three-dimensional structures of nucleic acids. Biophys. J. 1992, **63**, 751–759.

Bindewald, E. and Shapiro, B.A. RNA secondary structure prediction from sequence alignments using a network of k-nearest neighbor classifiers. RNA 2006;12(3):342-352.

Boden M, Yuan Z, Bailey T. Prediction of protein continuum secondary structure with probabilistic models based on NMR solved structures. BMC Bioinformatics. 2006; 7:68.

Boden M, Bailey TL. Identifying sequence regions undergoing conformational change via predicted continuum secondary structure. Bioinformatics. 2006; 22:1809–1814.

Bowie, J.U., Luthy, R. and Eisenberg, D. A method to identify protein sequences that fold into a known three-dimensional structure. *Science* 1991;253(5016):164-170.

Buck,A.H. *et al.* Structural perspective on the activation of RNase P RNA by protein. *Nat. Struc. Mol. Biol. 2005*, **12**, 958–964.

Cao, Song, and Shi-Jie Chen. "Predicting RNA folding thermodynamics with a reduced chain representation model." RNA 2005, 11(12): 1884-1897.

Caprara,A. et al. 1001 optimal PDB structure alignments: integer programming methods for finding the maximum contact map overlap. J. Comput. Biol., 2004, 11, 27–52.

Chen, X.*, et al.* FlexStem: improving predictions of RNA secondary structures with pseudoknots by reducing the search space. *Bioinformatics* 2008, 24(18):1994-2001.

Cheng,J. and Baldi,P. Improved residue contact prediction using support vector machines and a large feature set. BMC Bioinformatics 2007, 8, 113.

Cuff, J.A. and Barton, G.J. Evaluation and improvement of multiple sequence methods for protein secondary structure prediction. Proteins*: Structure, Function, and Bioinformatics* 1999;34(4):508-519.

Das, R. and Baker, D. Automated de novo prediction of native-like RNA tertiary structures. *Proceedings of the National Academy of Sciences* 2007;104(37):14664-14669.

Das, R. *et al*. Atomic accuracy in predicting and designing noncanonical RNA structure. *Nat. Methods 2010*, **7**, 291–294.

DeBartolo J, Colubri A, Jha AK, Fitzgerald JE, et al. Mimicking the folding pathway to improve homology-free protein structure prediction. Proc Natl Acad Sci USA. 2009; 106:3734–3739.

Di Lena,P. et al. Deep architectures for protein contact map prediction. Bioinformatics 2012, 28, 2449–2457

Ding,F. *et al.* Ab initio RNA folding by discrete molecular dynamics: from structure prediction to folding mechanisms. *RNA 2008*, **14**, 1164–1173.

Do, C.B., Woods, D.A. and Batzoglou, S. CONTRAfold: RNA secondary structure prediction without physics-based models. Bioinformatics 2006;22(14):e90-e98.

Duan M, Huang M, Ma C, Li L, Zhou Y. Position-specific residue preference features around the ends of helices and strands and a novel strategy for the prediction of secondary structures. Protein Sci. 2008; 17:1505–1512.

Eddy, S.R. and Durbin, R. RNA sequence analysis using covariance models. *Nucleic acids research* 1994;22(11):2079-2088.

Fariselli,P. and Casadio,R. A neural network based predictor of residue contacts in proteins. Protein Eng. 1999, 12, 15–21.

Ferretti,V. and Sankoff,D. A continuous analog for RNA folding. *B. Math. Biol. 1989*, **51**, 167–171.

Flores,S.C. *et al.* Predicting RNA structure by multiple template homology modeling. In *Pacific Symposium on Biocomputing 2010*. World Scientific Publishing, Co., Hawaii, pp. 216–227.

Frellsen, J*., et al.* A probabilistic model of RNA conformational space. *PLoS computational biology* 2009;5(6):e1000406.

Gardner,P. and Giegerich,R. A comprehensive comparison of comparative RNA structure prediction approaches. *BMC Bioinformatics 2004*, **5**, 140.

Gardner,P.P. *et al.* Rfam: updates to the RNA families database. *Nucleic Acids Res. 2009*, **37**, 136.

Giddings, J.C. and Byring, H. A molecular dynamic theory of chromatography. *The Journal of Physical Chemistry* 1955;59(5):416-421.

Gillespie,J. *et al.* RNA folding on the 3D triangular lattice. *BMC Bioinformatics 2009*, **10**, 369.

Gobel, U. et al. Correlated mutations and residue contacts in proteins. Proteins 2004, 18, 309–317.

Guo J, Chen H, Sun Z, Lin Y. A novel method for protein secondary structure prediction using dual-layer SVM and profiles. Proteins. 2004; 54:738–743.

Hajdin,C.E. *et al.* On the significance of an RNA tertiary structure prediction. *RNA 2010*, **16**, 1340–1349.

Hamada,M. *et al.* Predictions of RNA secondary structure by combining homologous sequence information. *Bioinformatics 2009*, **25**, 330.

Haspel,N. *et al.* Reducing the computational complexity of protein folding via fragment folding and assembly. *Protein Sci. 2003*, **12**, 1177–1187.

Havgaard,J.H. *et al.* Fast pairwise structural RNA alignments by pruning of the dynamical programming matrix, *PLoS Comput. Biol. 2007*, **3**, e193.

Hiller,M. *et al.* Pre-mRNA secondary structures influence exon recognition. *PLoS Genet. 2007*, **3**, e204.

Hofacker, I.L. Vienna RNA secondary structure server. *Nucleic acids research* 2003;31(13):3429-3431.

Holley, L.H. and Karplus, M. Protein secondary structure prediction with a neural network. *Proceedings of the National Academy of Sciences* 1989;86(1):152-156.

Hua S, Sun Z. A novel method of protein secondary structure prediction with high segment overlap measure: support vector machine approach. J Mol Biol. 2001; 308:397–407.

Jones, D.T. Protein secondary structure prediction based on position-specific scoring matrices. *Journal of molecular biology* 1999;292(2):195-202.

Jones, D.T*., et al.* PSICOV: precise structural contact prediction using sparse inverse covariance estimation on large multiple sequence alignments. *Bioinformatics* 2012;28(2):184-190.

Jonikas,M.A. *et al.* Coarse-grained modeling of large RNA molecules with knowledge-based potentials and structural filters. *RNA 2009*, **15**, 189–199.

Jonikas, Magdalena A., Randall J. Radmer, and Russ B. Altman. "Knowledge-based instantiation of full atomic detail into coarse-grain RNA 3D structural models." Bioinformatics 25.24 (2009): 3259-3266.

Kabsch, W. and Sander, C. Dictionary of protein secondary structure: pattern recognition of hydrogen‐bonded and geometrical features. *Biopolymers* 1983;22(12):2577-2637.

Karplus, M. and Weaver, D.L. Protein folding dynamics: The diffusion‐collision model and experimental data. *Protein Science* 1994;3(4):650-668.

Kim H, Park H. Protein secondary structure prediction based on an improved support vector machines approach. Protein Eng. 2003; 16:553–560.

Klepeis, J. and Floudas, C. ASTRO-FOLD: a combinatorial and global optimization framework for ab initio prediction of three-dimensional structures of proteins from the amino acid sequence. *Biophysical Journal* 2003;85(4):2119-2146.

Kneller, D., Cohen, F. and Langridge, R. Improvements in protein secondary structure prediction by an enhanced neural network. *Journal of molecular biology* 1990;214(1):171-182.

Knudsen, B. and Hein, J. Pfold: RNA secondary structure prediction using stochastic context-free grammars. *Nucleic acids research* 2003;31(13):3423-3428.

Lafferty, JD.; McCallum, A.; Pereira, FCN. Proceeding of International Conference on Machine Learning. Morgan Kaufmann Publishers Inc; San Francisco, CA, USA: 2001. p. 282-289.

Laing,C. and Schlick,T. Computational approaches to 3D modeling of RNA. *J. Phys.: Condens. Matter 2010*, **22**, 283101.

Lee,J. *et al.* Prediction of protein tertiary structure using PROFESY, a novel method based on fragment assembly and conformational space annealing. *Proteins: Struct., Funct., Bioinformatics 2004*, **56**, 704–714.

*Liu DC, Nocedal J. On the limited memory BFGS method for large scale optimization. Math Progr. 1989; 45:503–528.*

Mathews,D.H. Revolutions in RNA secondary structure prediction. *J. Mol. Biol. 2006*, **359**, 526–532.

Mathews,D.H. and Turner,D.H. Dynalign: an algorithm for finding the secondary structure common to two RNA sequences. *J. Mol. Biol. 2002*, **317**, 191–203.

Mathews,D.H. and Turner,D.H. Prediction of RNA secondary structure by free energy minimization. *Curr. Opin. Struc. Biol. 2006*, **16**, 270–278.

Meiler J, Müller M, Zeidler A, Schmäschke F. Generation and evaluation of dimension-reduced amino acid parameter representations by artificial neural networks. J Mol Model. 2001; 7:360– 369.

Morcos, F.*, et al.* Direct-coupling analysis of residue coevolution captures native contacts across many protein families. *Proceedings of the National Academy of Sciences* 2011;108(49):E1293-E1301.

Myers, J.K. and Oas, T.G. Preorganized secondary structure as an important determinant of fast protein folding. *Nature Structural & Molecular Biology* 2001;8(6):552-558.

Olmea,O. and Valencia,A. Improving contact predictions by the combination of correlated mutations and other sources of sequence information. Fold. Des. 1997, 2, S25–S32.

Ortiz, A.R.*, et al.* Ab initio folding of proteins using restraints derived from evolutionary information. *Proteins: Structure, Function, and* Bioinformatics 1999;37(S3):177-185.

Parisien, M. and Major, F. The MC-Fold and MC-Sym pipeline infers RNA structure from sequence data. *Nature* 2008;452(7183):51-55.

Pauling, L., Corey, R.B. and Branson, H.R. The structure of proteins: two hydrogen-bonded helical configurations of the polypeptide chain. *Proceedings of the National Academy of Sciences* 1951;37(4):205-211.

Peng, J.; Bo, L.; Xu, J. Advances in Neural Information Processing Systems. Bengio, Y.; Schuurmans, D.; Lafferty, J.; Williams, CKI.; Culotta, A., editors. Vol. 22. MIT Press; 2009. p. 1419-1427.

Peng J, Xu J. Low-homology protein threading. Bioinformatics. 2010; 26:294.

Pirovano, W. and Heringa, J. Protein secondary structure prediction. In, *Data Mining Techniques for the Life Sciences*. Springer; 2010. p. 327-348.

Pollastri, G., *et al.* Improving the prediction of protein secondary structure in three and eight classes using recurrent neural networks and profiles. *Proteins: Structure, Function, and Bioinformatics* 2002;47(2):228-235.

Poolsap,U. *et al.* Prediction of RNA secondary structure with pseudoknots using integer programming. *BMC Bioinformatics 2009*, **10**, S38.

Punta,M. and Rost,B. PROFcon: novel prediction of long-range contacts. Bioinformatics 2005, 21, 2960–2968.

Qian, N. and Sejnowski, T.J. Predicting the secondary structure of globular proteins using neural network models. *Journal of* molecular *biology* 1988;202(4):865-884.

Rabiner LR. A tutorial on hidden Markov models and selected applications in speech recognition. Proceedings of the IEEE. 1989:275–286.

Ray,P.S. *et al.* A stress-responsive RNA switch regulates VEGFA expression. *Nature 2009*, **457**, 915–919.

Reymond,C. *et al.* Modulating RNA structure and catalysis: lessons from small cleaving ribozymes. *Cell. Mol. Life Sci. 2009*, **66**, 3937–3950.

Rost B. PHD: predicting one-dimensional protein structure by profile-based neural networks. Methods Enzymol. 1996; 266:525–539.

Rost, B. and Sander, C. Prediction of protein secondary structure at better than 70% accuracy. *Journal of molecular biology* 1993;232(2):584-599.

Rost B, Sander C, Schneider R. Redefining the goals of protein secondary structure prediction. J Mol Biol. 1994; 235:13–26.

Rost, B. and Sander, C. Combining evolutionary information and neural networks to predict protein secondary structure. *Proteins: Structure, Function, and Bioinformatics* 1994;19(1):55-72.

Sato,K. and Sakakibara,Y. RNA secondary structural alignment with conditional random fields. *Bioinformatics 2005*, **21**, ii237–ii242.

Schäffer, A.A.*, et al.* Improving the accuracy of PSI-BLAST protein database searches with composition-based statistics and other refinements. *Nucleic acids research* 2001;29(14):2994-3005.

Sharma,S. *et al.* iFoldRNA: three-dimensional RNA structure prediction and folding. *Bioinformatics 2008*, **24**, 1951–1952.

Shen, Minyi, and Andrej Sali. "Statistical potential for assessment and prediction of protein structures." Protein science 15.11 (2006): 2507-2524.

Simons,K.T. *et al.* Assembly of protein tertiary structures from fragments with similar local sequences using simulated annealing and bayesian scoring functions. *J. Mol. Biol. 1997*, **268**, 209–225.

Solnick,D. Alternative splicing caused by RNA secondary structure. *Cell 1985*, **43**, 667–676.

Subramaniam, S., Tcheng, D.K. and Fenton, J.M. A knowledge-based method for protein structure refinement and prediction. Proceedings of International

Conference on Intelligent Systems for Molecular Biology ; ISMB. International Conference on Intelligent Systems for Molecular Biology 1996;4:218-229.

Sutcliffe, M., *et al.* Knowledge based modelling of homologous proteins, Part I: Three-dimensional frameworks derived from the simultaneous superposition of multiple structures. *Protein Engineering* 1987;1(5):377-384.

Tan YH, Huang H, Kihara D. Statistical potential-based amino acid similarity matrices for aligning distantly related protein sequences. Proteins. 2006; 64:587–600.

Tang,X. *et al.* Using motion planning to study RNA folding kinetics. *J. Comput. Biol. 2005*, **12**, 862–881.

Tegge,A.N. et al. NNcon: improved protein contact map prediction using 2D-recursive neural networks. Nucleic Acids Res. 2009, 37, W515–W518

Vassura, M., *et al.* Reconstruction of 3D structures from protein contact maps. *IEEE/ACM Transactions on* Computational *Biology and Bioinformatics (TCBB)* 2008;5(3):357-367.

Vendruscolo, M., Kussell, E. and Domany, E. Recovery of protein structure from contact maps. *Folding and Design* 1997;2(5):295-306.

Vendruscolo,M. and Domany,E. Pairwise contact potentials are unsuitable for protein folding. J. Chem. Phys. 1998, 109, 11101.

Vullo,A. et al. A two-stage approach for improved prediction of residue contact maps. BMC Bioinformatics 2006, 7, 180.

Wang G, Dunbrack RLJ. PISCES: a protein sequence culling server. Bioinformatics. 2003; 19:1589–1591.

Wang, Z. and Xu, J. A conditional random fields method for RNA sequence–structure relationship modeling and conformation sampling. *Bioinformatics* 2011;27(13):i102-i110.

Wang, Z. and Xu, J. Predicting protein contact map using evolutionary and physical constraints by integer programming. *Bioinformatics* 2013;29(13):i266-i273.

Ward JJ, McGuffin LJ, Buxton BF, Jones DT. Secondary structure prediction with support vector machines. Bioinformatics. 2003; 19:1650–1655.

Wexler,Y. *et al*. A study of accessible motifs and RNA folding complexity. *Res. Comput. Mol. Biol. 2006*, **3909**, 473–487.

Will,S. *et al.* Inferring noncoding RNA families and classes by means of genome-scale structure-based clustering. *PLoS Comput. Biol. 2007*, **3**, e65.

Wu,S. and Zhang,Y. A comprehensive assessment of sequence-based and template-based methods for protein contact prediction. Bioinformatics 2008, 24, 924–931.

Wu, S., Szilagyi, A. and Zhang, Y. Improving protein structure prediction using multiple sequence-based contact predictions. *Structure* 2011;19(8):1182-1191.

Xu,J. et al. A parameterized algorithm for protein structure alignment. J. Comput. Biol., 2007, 14, 564–577.

Zhang,J. *et al.* Discrete state model and accurate estimation of loop entropy of RNA secondary structures. *J. Chem. Phys. 2008*, **128**, 125107.

Zhao F, Li S, Sterner BW, Xu J. Discriminative learning for protein conformation sampling. Proteins. 2008; 73:228–240.

Zhao, F.; Peng, J.; DeBartolo, J.; Freed, K., et al. Research in Computational Molecular Biology. Batzoglou, S., editor. Springer; Berlin/Heidelberg: 2009. p. 59-73.

Zhao,F. *et al*. Fragment-free approach to protein folding using conditional neural fields. *Bioinformatics 2010*, **26**, i310–i317.

Zhou,H. and Skolnick,J. Protein model quality assessment prediction by combining fragment comparisons and a consensus Cα contact potential. Proteins, 2007, 71, 1211–1218.

Zuker, M. Mfold web server for nucleic acid folding and hybridization prediction. *Nucleic acids research* 2003;31(13):3406-3415.

Zuker, M. and Sankoff, D. RNA secondary structures and their prediction. *Bulletin of Mathematical Biology* 1984;46(4):591-621.